\documentclass[12pt]{iopart}
\renewcommand{\today}{To be published in {\sl New Journal of Physics}, 
September 2014}
\usepackage{graphicx}

\newcommand{\vecr}{{\bf r}}
\newcommand{\vecR}{{\bf R}}
\newcommand{\veck}{{\bf k}}

\newcommand{\vecG}{{\bf G}}

\newcommand{\llambda}{\kappa}

\begin{document}

\title[Gutzwiller Density Functional Theory]%
{Gutzwiller Density Functional Theory: 
a formal derivation and application to ferromagnetic nickel}

\author{Tobias Schickling$^{1}$, J\"org B\"unemann$^{1}$, Florian Gebhard$^{1,2}$, 
and Werner Weber$^{3,\dagger}$}

\address{$^{1}$ Fachbereich Physik, Philipps Universit\"at Marburg, D-35032
Marburg, Germany}

\address{$^{2}$ Material Sciences Center, Philipps Universit\"at Marburg, D-35032
Marburg, Germany}

\address{$^{3}$ Institut f\"ur Physik, Technische Universit\"at Dortmund, 
D-44227 Dortmund, Germany}

\footnotetext{deceased on July 3, 2014}
\begin{abstract}%
We  present a detailed derivation of the Gutzwiller 
Density Functional Theory that 
covers all conceivable cases of symmetries and 
Gutzwiller wave functions. The method is used in a  
study of ferromagnetic nickel where we calculate 
ground state properties (lattice constant, bulk modulus, spin magnetic moment) 
and the quasi-particle band structure. Our method resolves
most shortcomings of an ordinary Density Functional
calculation on nickel. However, the quality of the results 
strongly depends on the particular choice of the double-counting correction. This 
constitutes a serious problem for all methods that attempt 
to merge Density Functional Theory with correlated-electron approaches 
based on Hubbard-type local interactions.     
\end{abstract}

\today
\maketitle

\section{Introduction}

Density Functional Theory (DFT) is the workhorse of 
electronic structure theory~\cite{MartinLDAgeneral}.
Based on the Hohenberg-Kohn theorem~\cite{MartinLDAgeneral},
the ground-state properties of an inter\-acting many-electron system
are calculated from those of an effective single-particle problem that can be solved
numerically. 
An essential ingredient in DFT is the so-called exchange-correlation potential
which, however, is unknown and sensible approximations must be devised,
e.g., the local (spin) density approximation, L(S)DA. 
In this way, the electronic properties of metals were calculated 
systematically~\cite{Moruzzi}. 
Unfortunately, the L(S)DA leads to unsatisfactory results 
for transition metals, their
compounds, and for heavy-fermion systems.
The electrons in the narrow $3d$ or $4f/5f$ bands experience correlations
that are not covered by current exchange-correlation potentials. 

For a more accurate description of electronic correlations in narrow bands,
Hubbard-type models~\cite{Hubbard,Gutzwiller} have been put forward.
However, simplistic model Hamiltonians can describe limited aspects
of real materials at best, while, at the same time,
they reintroduce the full complexity of the many-body problem.
Recently, new methods were developed that permit
the (numerical) analysis of multi-band Hubbard models,
and, moreover, can be combined with DFT, 
specifically, the LDA+$U$ method~\cite{Lichtenstein}, 
the LDA+DMFT (Dynamical Mean-Field 
Theory)~\cite{LDA-DMFT-kotliar,LDA-DMFT-vollhardt},
and the Gutzwiller variational 
approach~\cite{buenemann2005,GutzwillerDFT1,GutzwillerDFT2,dong2014}.
The LDA+$U$ approach treats atomic interactions on a mean-field level
so that it is computationally cheap but it ignores true many-body correlations.
The DMFT becomes formally exact 
for infinite lattice coordination number, $Z\to \infty$, but it requires
the self-consistent solution of a dynamical impurity problem 
that is numerically very demanding.
The Gutzwiller DFT is based on a variational
treatment of local many-body correlations. Expectation values
can be calculated for $Z\to \infty$ without further approximations,
and the remaining computational problem remains tractable.

Previously, we used the DFT to obtain the bare band 
structure from which we calculated the properties
of nickel~\cite{buenemann2005,nickelpaper1,nickelpaper2,nickelpaper3}  
and LaOFeAs~\cite{LaOFeAs}. 
For these studies, we developed a formalism that applies to
general Gutzwiller-correlated states for arbitrary multi-band Hubbard Hamiltonians.  
However, some single-particle properties remained fixed at their DFT values.
In contrast, in Refs.~\cite{GutzwillerDFT1,GutzwillerDFT2,dong2014}
the correlated electron density was fed back into the DFT calculations
but the Gutzwiller quasi-particle Hamiltonian was introduced in an ad-hoc manner. 
In this work, 
%
%
%
%
we present a formal derivation of the Gutzwiller DFT 
as a generic extension of the DFT.
Our formulae apply for general Gutzwiller-correlated wave functions
and reproduce expressions used 
previously~\cite{GutzwillerDFT1,GutzwillerDFT2} as special
cases; 
%
%
for a recent application to topological insulators, 
see Ref.~\cite{Chinesentruppe}.
%
%
Here, we give results for nickel in face-centered cubic
structure. The Gutzwiller DFT results
for the lattice constant, the magnetic spin-only moment, and 
the bulk modulus agree very well with experiments. 
Moreover, the quasi-particle bandstructure from Gutzwiller DFT 
is in satisfactory agreement with data from Angle-Resolved Photo-Emission Spectroscopy
(ARPES). As found earlier~\cite{GutzwillerDFT1,GutzwillerDFT2}, the Gutzwiller DFT
overcomes the limitations of DFT for the description of
transition metals.

Our paper is organized as follows.
In Sect.~\ref{sec:DFT} we recall the derivation of Density Functional Theory (DFT)
as a variational approach to the many-body problem
and its mapping to an effective single-particle reference system
(Kohn-Sham scheme).
We extend our concise derivation
to many-particle reference systems
in Sect.~\ref{sec:GDFT}. In particular, we formulate the Gutzwiller density functional
whose minimization leads to the Gutzwiller--Kohn-Sham Hamiltonian.
The theory is worked out in the limit of large coordination number, $Z\to \infty$,
where explicit expressions for the Gutzwiller density functional are available.
In Sect.~\ref{sec:transinvsystems} we restrict ourselves to 
lattice systems that are invariant under translation by a lattice vector
so that the quasi-particle excitations can be characterized 
by their Bloch momentum. In Sect.~\ref{sec:resultsnickel} we present results 
for face-centered cubic (fcc) nickel where $Z=12$.
A short summary, Sect.~\ref{sec:summary}, closes our presentation.
Technical details are deferred to the appendices.

\section{Density Functional Theory}
\label{sec:DFT}

We start our presentation with a concise derivation of
Density Functional Theory that can readily be extended 
to the Gutzwiller Density Functional Theory.

\subsection{Many-particle Hamiltonian and Ritz variational principle}

Our many-particle Hamiltonian for electrons with spin 
$\sigma=\uparrow,\downarrow$ reads ($\hbar \equiv 1$)
\begin{eqnarray}
\hat{H}&=&\hat{H}_{\rm band} + \hat{H}_{\rm int}\nonumber \; ,\\
\hat{H}_{\rm band} &=& \sum_{\sigma} \int {\rm d}\vecr 
\hat{\Psi}_{\sigma}^{\dagger}(\vecr)
\left(
-\frac{\Delta_{\vecr}}{2m} 
+U(\vecr) 
\right)
\hat{\Psi}_{\sigma}^{\vphantom{\dagger}}(\vecr) \; , \nonumber\\
\hat{H}_{\rm int}&=& \sum_{\sigma,\sigma'} 
\int {\rm d}\vecr \int {\rm d}\vecr' 
\hat{\Psi}_{\sigma}^{\dagger}(\vecr) \hat{\Psi}_{\sigma'}^{\dagger}(\vecr')V(\vecr-\vecr')
\hat{\Psi}_{\sigma'}^{\vphantom{\dagger}}(\vecr') 
\hat{\Psi}_{\sigma}^{\vphantom{\dagger}}(\vecr)
\label{eq:Hamiltondef}
\end{eqnarray}
with 
\begin{equation}
V(\vecr-\vecr')=\frac{1}{2}\frac{e^2}{|\vecr-\vecr'|} \; .
\end{equation}
The electrons experience the periodic potential of the ions, $U(\vecr)$,
and their mutual Coulomb interaction, $V(\vecr-\vecr')$.
The total number of electrons is $N=N_{\uparrow}+N_{\downarrow}$.
According to the Ritz variational principle, the ground
state of a Hamiltonian $\hat{H}$ can be obtained from
the minimization of the energy functional
\begin{equation}
E\left[\left\{|\Psi\rangle\right\}\right]
= \langle \Psi | \hat{H} | \Psi \rangle 
\label{eq:Ritz}
\end{equation}
in the subset of normalized states $|\Psi\rangle$ in the Hilbert space
 with $N$ electrons, $\langle \Psi | \Psi \rangle =1$.

\subsection{Levy's constrained search}
\label{subsec:Levy}

The minimization of the energy functional~(\ref{eq:Ritz})
is done in two steps, the constrained search~\cite{Levy}, Sect.~\ref{subsubsec:CS},
and the minimization of the density functional, Sect.~\ref{subsubsec:DFmini}.
To this end, we consider the subset of normalized states $|\Psi^{(n)}\rangle$
with fixed electron densities~$n_{\sigma}(\vecr)$,
\begin{equation}
n_{\sigma}(\vecr)= \langle \Psi^{(n)} | \hat{\Psi}_{\sigma}^{\dagger}(\vecr)
\hat{\Psi}_{\sigma}^{\vphantom{\dagger}}(\vecr) | \Psi^{(n)} \rangle \; .
\label{eq:electrondensity}
\end{equation}
In the following we accept `physical' densities only, i.e.,
those $n_{\sigma}(\vecr)$ for which states $| \Psi^{(n)} \rangle$ can be found.
For the subset of states $| \Psi^{(n)} \rangle$ we define 
\begin{eqnarray}
\hat{H}_{\rm e}&=&\hat{H}_{\rm kin} + \hat{V}_{\rm xc}\; ,\\
\hat{H}_{\rm kin} &=& \sum_{\sigma} \int {\rm d}\vecr 
\hat{\Psi}_{\sigma}^{\dagger}(\vecr)  \left(-\frac{\Delta_{\vecr}}{2m} \right)
\hat{\Psi}_{\sigma}^{\vphantom{\dagger}}(\vecr)\; ,
\label{eq:Hkinalone} 
\\
\hat{V}_{\rm xc}&=& \sum_{\sigma,\sigma'} 
\int {\rm d}\vecr \int {\rm d}\vecr' V(\vecr-\vecr')
\left[
\hat{\Psi}_{\sigma}^{\dagger}(\vecr) \hat{\Psi}_{\sigma'}^{\dagger}(\vecr')
\hat{\Psi}_{\sigma'}^{\vphantom{\dagger}}(\vecr') 
\hat{\Psi}_{\sigma}^{\vphantom{\dagger}}(\vecr)
\right. 
\nonumber \\
&& 
\left.
-
\hat{\Psi}_{\sigma}^{\dagger}(\vecr) \hat{\Psi}_{\sigma}^{\vphantom{\dagger}}(\vecr)
n_{\sigma'}(\vecr') 
-
\hat{\Psi}_{\sigma'}^{\dagger}(\vecr')\hat{\Psi}_{\sigma'}^{\vphantom{\dagger}}(\vecr') 
n_{\sigma}(\vecr)
+
n_{\sigma}(\vecr)n_{\sigma'}(\vecr')
\right] \; .
\label{eq:HkinVxc}
\end{eqnarray}
Here, we extracted the Hartree terms from
the Coulomb interaction $H_{\rm int}$ in eq.~(\ref{eq:Hamiltondef})
so that $\hat{V}_{\rm xc}$ contains
only the so-called exchange and correlation contributions.
In the subset of normalized states $|\Psi^{(n)}\rangle$ 
we consider the functional
\begin{equation}
F\left[\left\{ n_{\sigma}(\vecr) \right\}, 
\left\{|\Psi^{(n)}\rangle\right\}\right]
= \langle \Psi^{(n)} | \hat{H}_{\rm e} | \Psi^{(n)} \rangle  \; .
\label{eq:Ritzreduced}
\end{equation}
For fixed densities $n_{\sigma}(\vecr)$,
the Hamiltonian $\hat{H}_{\rm e}$ defines an electronic 
problem where the periodic potential of the ions is formally absent.

\subsubsection{Constrained search.}
\label{subsubsec:CS}

The formal task is to find the minimum 
of the energy functional $F$ in~(\ref{eq:Ritzreduced}) with 
respect to $|\Psi^{(n)}\rangle$,
\begin{equation}
\bar{F}\left[\left\{ n_{\sigma}(\vecr) \right\}\right] 
={\rm Min}_{\left\{|\Psi^{(n)}\rangle\right\}}
F\left[\left\{ n_{\sigma}(\vecr) \right\}, 
\left\{|\Psi^{(n)}\rangle\right\}\right]\; .
\label{eq:defF-functional}
\end{equation}
Recall that the electron densities $n_{\sigma}(\vecr)$ are fixed
in this step. 
We denote the resulting optimal many-particle state $|\Psi_0^{(n)}\rangle$.
Thus, we may write 
\begin{equation}
\bar{F}\left[\left\{ n_{\sigma}(\vecr) \right\}\right] 
=F \left[\left\{ n_{\sigma}(\vecr) \right\}, 
\left\{|\Psi_0^{(n)}\rangle\right\}\right]
= \langle \Psi_0^{(n)} | \hat{H}_{\rm e} | \Psi_0^{(n)} \rangle  \; .
\label{eq:defF-functionalPsizero}
\end{equation}
For later use, we define the functionals for the kinetic energy
\begin{equation}
K\left[\left\{ n_{\sigma}(\vecr) \right\}\right] 
=\langle \Psi_0^{(n)} | \hat{H}_{\rm kin}| \Psi_0^{(n)} \rangle 
\label{eq:defK-functional}
\end{equation}
and the exchange-correlation energy
\begin{equation}
E_{\rm xc}\left[\left\{ n_{\sigma}(\vecr) \right\}\right] 
=\langle \Psi_0^{(n)} | \hat{V}_{\rm xc}| \Psi_0^{(n)} \rangle
\label{eq:defExc-functional}
\end{equation}
so that
\begin{equation}
\bar{F}\left[\left\{ n_{\sigma}(\vecr) \right\}\right] 
= K\left[\left\{ n_{\sigma}(\vecr) \right\}\right]  +
E_{\rm xc}\left[\left\{ n_{\sigma}(\vecr) \right\}\right] \; .
\label{eq:deftotal-functional}
\end{equation}

\subsubsection{Density functional, ground-state density 
and ground-state energy.}
\label{subsubsec:DFmini}

After the constrained search as a first step, 
we are left with the density functional
\begin{eqnarray}
D\left[\left\{ n_{\sigma}(\vecr) \right\}\right] &=&
\bar{F}\left[\left\{ n_{\sigma}(\vecr) \right\}\right] 
+ U\left[\left\{ n_{\sigma}(\vecr) \right\}\right] 
+ V_{\rm Har}\left[\left\{ n_{\sigma}(\vecr) \right\}\right]
\label{eq:dintparticles}\\
&=& 
K\left[\left\{ n_{\sigma}(\vecr) \right\}\right] 
+ U\left[\left\{ n_{\sigma}(\vecr) \right\}\right] 
+ V_{\rm Har}\left[\left\{ n_{\sigma}(\vecr) \right\}\right] 
+ E_{\rm xc}\left[\left\{ n_{\sigma}(\vecr) \right\}\right] 
\nonumber 
\end{eqnarray}
with
\begin{eqnarray}
U\left[\left\{ n_{\sigma}(\vecr) \right\}\right] 
&=&  \sum_{\sigma}\int {\rm d}\vecr 
U(\vecr) n_{\sigma}(\vecr)  \; , \nonumber \\
V_{\rm Har}\left[\left\{ n_{\sigma}(\vecr) \right\}\right] 
&=& 
\sum_{\sigma,\sigma'} \int {\rm d}\vecr \int {\rm d}\vecr' 
V(\vecr-\vecr') n_{\sigma}(\vecr) n_{\sigma'}(\vecr') \; .
\label{eq:UandVH}
\end{eqnarray}

According to the Ritz variational principle,
the ground-state energy $E_0$ 
is found from the minimization of this functional
over the densities $n_{\sigma}(\vecr)$,
\begin{equation}
E_0={\rm Min}_{\left\{ n_{\sigma}(\vecr) \right\}}
D\left[\left\{ n_{\sigma}(\vecr) \right\}\right] \; .
\end{equation}
The ground-state densities $n_{\sigma}^0(\vecr)$ are those where
the minimum of $D\left[\left\{ n_{\sigma}(\vecr) \right\}\right]$
is obtained.

\subsection{Single-particle reference system}
\label{subsec:SPHamiltonian}

We consider the subset of single-particle
product states $|\Phi^{(n)}\rangle$ that are normalized to unity,
$\langle \Phi^{(n)} | \Phi^{(n)} \rangle=1$. As before, the upper index indicates that
they all lead to the same (physical) 
single-particle densities~$n_{\sigma}^{\rm sp}(\vecr)$,
\begin{equation}
n_{\sigma}^{\rm sp}(\vecr) 
= \langle \Phi^{(n)}  | \hat{\Psi}_{\sigma}^{\dagger}(\vecr)
\hat{\Psi}_{\sigma}(\vecr)^{\vphantom{\dagger}} | \Phi^{(n)}  \rangle \; .
\label{eq:defdensityagain}
\end{equation}
As our single-particle Hamiltonian we consider 
the kinetic-energy operator~$\hat{H}_{\rm kin}$, see
eq.~(\ref{eq:Hkinalone}). 
For fixed single-particle densities~$n_{\sigma}^{\rm sp}(\vecr)$
we define the single-particle kinetic-energy functional
\begin{equation}
F_{\rm sp}\left[\left\{ n_{\sigma}^{\rm sp}(\vecr) \right\},
\left\{ |\Phi^{(n)}\rangle\right\}
\right]=
\langle \Phi^{(n)} | \hat{H}_{\rm kin} | \Phi^{(n)} \rangle 
\; .
\end{equation}

\subsubsection{Constrained search.}
\label{subsec:Fsp}

As in Sect.~\ref{subsec:Levy}, we carry out a constrained search
in the subset of states $|\Phi^{(n)}\rangle$.
The task is the minimization of the kinetic-energy functional
$F_{\rm sp}\left[\left\{ n_{\sigma}^{\rm sp}(\vecr) \right\},\left\{ |\Phi^{(n)}\rangle\right\}
\right]$.
We denote the optimized single-particle product state $|\Phi_0^{(n)}\rangle$
so that we find the density functional for the kinetic energy as
\begin{equation}
\bar{F}_{\rm sp}\left[\left\{ n_{\sigma}^{\rm sp}(\vecr) \right\}\right]=
\langle \Phi_0^{(n)} | \hat{H}_{\rm kin} | \Phi_0^{(n)} \rangle 
\equiv K_{\rm sp}\left[\left\{ n_{\sigma}^{\rm sp}(\vecr) \right\}\right]
\; .
\label{eq:Kkin}
\end{equation}

\subsubsection{Single-particle density functional.}
\label{subsubsec:DFminiSP}

As the density functional 
$D_{\rm sp}\left[\left\{ n_{\sigma}^{\rm sp}(\vecr) \right\}\right] $
that corresponds to the single-particle problem we define
\begin{eqnarray}
D_{\rm sp}\left[\left\{ n_{\sigma}^{\rm sp}(\vecr) \right\}\right] &=&
K_{\rm sp}\left[\left\{ n_{\sigma}^{\rm sp}(\vecr) \right\}\right] 
+ U\left[\left\{ n_{\sigma}^{\rm sp}(\vecr) \right\}\right] 
+ V_{\rm Har}\left[\left\{ n_{\sigma}^{\rm sp}(\vecr) \right\}\right]  
\nonumber \\
&& +E_{\rm sp,xc}\left[\left\{ n_{\sigma}^{\rm sp}(\vecr) \right\}\right] \; ,
\label{eq:Dsp}
\label{eq:denssp} 
\end{eqnarray}
with the kinetic energy term from~(\ref{eq:Kkin}),
the contributions from the external potential and the Hartree terms 
$U\left[\left\{ n_{\sigma}^{\rm sp}(\vecr) \right\}\right]$
and $V_{\rm Har}\left[\left\{ n_{\sigma}^{\rm sp}(\vecr) \right\}\right]$ from 
eq.~(\ref{eq:UandVH}),
and the single-particle exchange-correlation potential
$E_{\rm sp, xc} \left[\left\{ n_{\sigma}^{\rm sp}(\vecr) \right\}\right]$
that we will specify later.
The functional~(\ref{eq:denssp}) defines our single-particle reference system.

\subsubsection{Noninteracting $V$-representability.}
\label{subsubsec:Vrepresentability}

In order to link the many-particle and single-particle approaches 
we make the assumption
of non-interacting $V$-representability~\cite{MartinLDAgeneral}:
For any given (physical) densities $n_{\sigma}(\vecr)$ we can find
a subset of normalized single-particle product states 
$|\Phi^{(n)}\rangle$ with $N$ electrons
such that 
\begin{equation}
n_{\sigma}^{\rm sp}(\vecr)=n_{\sigma}(\vecr) \; .
\end{equation}
Moreover, we demand that the density functionals
$D\left[\left\{ n_{\sigma}(\vecr) \right\}\right]$~(\ref{eq:dintparticles})
for the interacting electrons 
and 
$D_{\rm sp}\left[\left\{ n_{\sigma}(\vecr) \right\}\right]$~(\ref{eq:denssp}) 
for the single-particle problem
agree with each other~\cite{note1},
\begin{equation}
D_{\rm sp}\left[\left\{ n_{\sigma}(\vecr) \right\}\right]=
D\left[\left\{ n_{\sigma}(\vecr) \right\}\right]  \; .
\label{eq:DisDsp}
\end{equation}
Then, the single-particle problem
leads to the same ground-state density $n_{\sigma}^{0}(\vecr)$
and ground-state energy~$E_0$
as the interacting-particle Hamiltonian 
because the density variation is done with the same density functional
(Hohenberg-Kohn theorem)~\cite{MartinLDAgeneral}.

The condition~(\ref{eq:DisDsp}) is equivalent to
\begin{equation}
K_{\rm sp}\left[\left\{ n_{\sigma}(\vecr) \right\}\right] 
+E_{\rm sp,xc}\left[\left\{ n_{\sigma}(\vecr) \right\}\right]
=
K\left[\left\{ n_{\sigma}(\vecr) \right\}\right] 
+E_{\rm xc}\left[\left\{ n_{\sigma}(\vecr) \right\}\right] 
\label{eq:FisFsp}
\end{equation}
because the interaction with the external potential and
the Hartree term only depend on the densities.
Eq.~(\ref{eq:FisFsp}) then leads to an exact expression
for the single-particle exchange-correlation energy
\begin{eqnarray}
E_{\rm sp, xc}\left[\left\{ n_{\sigma}(\vecr) \right\}\right] 
&=& K\left[\left\{ n_{\sigma}(\vecr) \right\}\right] 
 -K_{\rm sp}\left[\left\{ n_{\sigma}(\vecr) \right\}\right] 
+E_{\rm xc}\left[\left\{ n_{\sigma}(\vecr) \right\}\right] \; . 
\label{eq:defExcsp}
\end{eqnarray}
This is our defining equation for 
$E_{\rm sp, xc}\left[\left\{ n_{\sigma}(\vecr) \right\}\right]$ in eq.~(\ref{eq:Dsp}).

\subsection{Kohn-Sham Hamiltonian}
\label{subsubsec:Kohn-Sham}

In the following we address the single-particle energy functional directly,
i.e., the Ritz variational problem without a prior constrained search,
\begin{eqnarray}
E\left[\left\{n_{\sigma}(\vecr)\right\},\left\{ |\Phi\rangle \right\}\right]
&=&\langle \Phi | \hat{H}_{\rm kin}  | \Phi\rangle
+U\left[\left\{n_{\sigma}(\vecr)\right\}\right]
+V_{\rm Har}\left[\left\{n_{\sigma}(\vecr)\right\}\right]  \nonumber \\
&& +E_{\rm sp,xc}\left[\left\{n_{\sigma}(\vecr)\right\}\right]
\; .
\end{eqnarray}
For the extension to the Gutzwiller Density Functional
Theory in Sect.~\ref{sec:GDFT}, we expand the field operators in a basis,
\begin{equation}
\hat{\Psi}_{\sigma}^{\vphantom{\dagger}}(\vecr)=\sum_{i} 
\langle \vecr | i, \sigma\rangle 
\hat{c}_{i,\sigma}^{\vphantom{\dagger}}
\quad , \quad
\hat{\Psi}_{\sigma}^{\dagger}(\vecr) =\sum_{i} 
\hat{c}_{i,\sigma}^{\dagger} 
\langle i, \sigma | \vecr \rangle
\; ,
\label{eq:expandpsi}
\end{equation}
where the index $i$ represents a combination
of site (or crystal momentum) index and an orbital index. 
For a canonical basis we must have
completeness and orthogonality,
\begin{equation}
\sum_{i,\sigma} |i,\sigma \rangle \langle i,\sigma| = \hat{1} \quad , \quad
\langle i,\sigma| j, \sigma'\rangle 
= \delta_{i,j} \delta_{\sigma,\sigma'}\; .
\label{eq:orthocomplete}
\end{equation}
When we insert~(\ref{eq:expandpsi}) into~(\ref{eq:Hkinalone}),
we obtain the operator for the kinetic energy 
in a general single-particle basis,
\begin{equation}
\hat{H}_{\rm kin}=
\sum_{i,j,\sigma} T_{i,j;\sigma}
\hat{c}_{i,\sigma}^{\dagger}\hat{c}_{j,\sigma}^{\vphantom{\dagger}} \;,
\label{eq:Kingeneralbasis}
\end{equation}
where the elements of the kinetic-energy matrix 
$\widetilde{T}_{\sigma}$ are given by
\begin{equation}
T_{i,j;\sigma}
%
= \int {\rm d}\vecr \xi_{i,\sigma}^*(\vecr) 
\left(-\frac{\Delta_{\vecr}}{2m} \right)
\xi_{j,\sigma}(\vecr)
\; ,
\label{eq:defineTij} 
\end{equation}
with $ \xi_{i,\sigma}(\vecr)=\langle \vecr | i,\sigma \rangle $. 

\subsubsection{Energy functional.}
We introduce the single-particle
density matrix $\tilde{\rho}$. Its elements in the general
single-particle basis read
\begin{equation}
\rho_{(i,\sigma),(j,\sigma)}
=\langle \Phi | \hat{c}_{j,\sigma}^{\dagger} \hat{c}_{i,\sigma}^{\vphantom{\dagger}}
| \Phi \rangle \equiv \rho_{i,j;\sigma}\; .
\label{eq:rhoelements}
\end{equation}
Then, the densities are given by
\begin{equation}
n_{\sigma}(\vecr)= \sum_{i,j} \xi_{i,\sigma}^*(\vecr)\xi_{j,\sigma}(\vecr)\rho_{j,i;\sigma} \; .
\label{eq:densitycond}
\end{equation}
Using these definitions, we can write the energy functional 
in the form
\begin{eqnarray}
E\left[\left\{n_{\sigma}(\vecr)\right\},\tilde{\rho}\right]
&=&\sum_{i,j}\sum_{\sigma} T_{i,j;\sigma}
\rho_{j,i;\sigma} +U\left[\left\{n_{\sigma}(\vecr)\right\}\right]
+V_{\rm Har}\left[\left\{n_{\sigma}(\vecr)\right\}\right] \nonumber\\
&&+E_{\rm sp,xc}\left[\left\{n_{\sigma}(\vecr)\right\}\right]
\; .
\label{eq:getmeEsp}
\end{eqnarray}
The fact that $|\Phi\rangle$ are normalized single-particle product states
is encoded in the matrix relation
\begin{equation}
\tilde{\rho}\cdot \tilde{\rho}
=\tilde{\rho} \; .
\label{eq:rhorho}
\end{equation}
This is readily proven by using a unitary transformation between the
operators $\hat{c}_{i,\sigma}$ and the single-particle operators
$\hat{b}_{k,\sigma}$ that generate $|\Phi\rangle$, 
see~\ref{app:showrhorhoequalsrho}.

When we minimize $E\left[\left\{n_{\sigma}(\vecr)\right\},\tilde{\rho}\right]$
with respect to $\tilde{\rho}$ we must take the condition~(\ref{eq:rhorho})
into account using a matrix $\widetilde{\Omega}$ of Lagrange multipliers 
$\Omega_{l,m;\sigma}$. Moreover, we
use the Lagrange multipliers $\llambda_{\sigma}(\vecr)$ to ensure
eq.~(\ref{eq:densitycond}), i.e., altogether we
address the functional $G_{\rm DFT}\equiv G_{\rm DFT}
\left[ \tilde{\rho} ,\widetilde{\Omega}, \left\{n_{\sigma}(\vecr)\right\},
\left\{ \llambda_{\sigma}(\vecr) \right\}\right] $
\begin{eqnarray}
G_{\rm DFT}&=&
E\left[\left\{n_{\sigma}(\vecr)\right\},\tilde{\rho}\right]
- \sum_{l,m,\sigma}\Omega_{l,m;\sigma} 
\biggl( \sum_p \rho_{l,p;\sigma}\rho_{p,m;\sigma}-\rho_{l,m;\sigma}
\biggr)\label{eq:minimizewithlambda}
\\
&&-
\sum_{\sigma} \int{\rm d}\vecr \llambda_{\sigma}(\vecr) 
\biggl(n_{\sigma}(\vecr)- \sum_{i,j} 
\xi_{i,\sigma}^*(\vecr)\xi_{j,\sigma}(\vecr)
\rho_{j,i;\sigma}\biggr)  \, .
\nonumber
\end{eqnarray}

\subsubsection{Minimization.}
When we minimize $G_{\rm DFT}$ in eq.~(\ref{eq:minimizewithlambda})
with respect to $n_{\sigma}(\vecr)$ we find 
{\arraycolsep=0.5pt\begin{eqnarray}
\llambda_{\sigma}(\vecr)&=& U(\vecr) + V_{\rm Har}(\vecr)
+v_{{\rm sp,xc},\sigma}(\vecr)
 \label{eq:lambdaissppotential}
\; ,\\
V_{\rm Har}(\vecr) &\equiv& \sum_{\sigma'}\int {\rm d}\vecr' 2V(\vecr-\vecr')
n_{\sigma'}^0(\vecr')
\; ,\\
v_{{\rm sp,xc},\sigma}(\vecr)&\equiv& 
\left.
\frac{\partial E_{\rm sp,xc}\left[\left\{ n_{\sigma'}(\vecr') \right\}\right]}%
{\partial n_{\sigma}(\vecr)} 
\right|_{n_{\sigma}(\vecr)=n_{\sigma}^0(\vecr)}
\label{eq:getmeexactKS}\\
&=& 
\left.\frac{\partial
\left[
K\left[\left\{ n_{\sigma'}(\vecr') \right\}\right] 
 -K_{\rm sp}\left[\left\{ n_{\sigma'}(\vecr') \right\}\right] 
+E_{\rm xc}\left[\left\{ n_{\sigma'}(\vecr') \right\}\right] 
\right]
}{\partial n_{\sigma}(\vecr)}
\right|_{n_{\sigma}(\vecr)=n_{\sigma}^0(\vecr)}
\nonumber \!\!, 
\end{eqnarray}}%
where $V_{\rm Har}(\vecr)$ is the Hartree interaction and $v_{{\rm sp,xc},\sigma}(\vecr)$
is the single-particle exchange-correlation potential.

The minimization with respect to $\tilde{\rho}$ is outlined 
in~\ref{appbueni}~\cite{lecturenotes}.
It leads to the Kohn-Sham single-particle Hamiltonian
\begin{equation}
\hat{H}^{\rm KS}=
\sum_{i,j,\sigma} T_{i,j;\sigma}^{\rm KS} 
\hat{c}_{i,\sigma}^{\dagger}\hat{c}_{j,\sigma}^{\vphantom{\dagger}} \;,
\label{eq:KSgeneralbasis}
\end{equation}
where the elements of the Kohn-Sham Hamilton matrix 
$\widetilde{T}_{\sigma}^{\rm KS}$ are given by
\begin{equation}
T_{i,j;\sigma}^{\rm KS} 
= \frac{\partial E\left[\left\{n_{\sigma}(\vecr)\right\},\tilde{\rho}\right]
}{
\partial \rho_{j,i;\sigma}
} 
+\int {\rm d}\vecr \llambda_{\sigma}(\vecr) 
\xi_{i,\sigma}^*(\vecr)\xi_{j,\sigma}(\vecr)
\; . \label{eq:TijKohnShamforever}
\end{equation}
Explicitly,
\begin{eqnarray}
T_{i,j;\sigma}^{\rm KS}
&=& 
\int {\rm d}\vecr \xi_{i,\sigma}^*(\vecr) 
h_{\sigma}^{\rm KS}(\vecr) 
\xi_{j,\sigma}(\vecr)
\label{eq:defineTijKS} 
\; ,\\
h_{\sigma}^{\rm KS}(\vecr) &\equiv & -\frac{\Delta_{\vecr}}{2m}
+ V_{\sigma}^{\rm KS}(\vecr) \; ,
\label{eq:definehKS} 
\\
V_{\sigma}^{\rm KS}(\vecr) &\equiv & \llambda_{\sigma}(\vecr) 
= U(\vecr) + V_{\rm Har}(\vecr) +v_{{\rm sp,xc},\sigma}(\vecr) \; .
\label{eq:defineVKSsss} 
\end{eqnarray}
Here, we defined the `Kohn-Sham potential' $V_{\sigma}^{\rm KS}(\vecr)$
that, in our derivation, is identical to the 
Lagrange parameter $\llambda_{\sigma}(\vecr)$.

The remaining task is to find the basis in which the Kohn-Sham 
matrix $\widetilde{T}_{\sigma}^{\rm KS}$ is diagonal, see \ref{app:basissets}.

\section{Density Functional Theory for many-particle reference systems}
\label{sec:GDFT}

The Kohn-Sham potential~(\ref{eq:getmeexactKS}) cannot be 
calculated exactly because the functionals in
eq.~(\ref{eq:defExcsp}) are not known. 
Therefore, assumptions must be made about the form
of the single-particle exchange-correlation potential, e.g., 
the Local Density Approximation~\cite{MartinLDAgeneral}.
Unfortunately, such approximations are not satisfactory for, e.g.,
transition metals and their compounds, and more sophisticated
many-electron approaches must be employed.

\subsection{Hubbard Hamiltonian and Hubbard density functional}
\label{subsubsec:Hubbardmodel}

\subsubsection{Multi-band Hubbard model.}

A better description of transition metals and their compounds
can be achieved by supplementing the single-particle reference system resulting from
$\hat{H}_{\rm kin}$ in Sect.~\ref{subsec:SPHamiltonian} by
a multi-band Hubbard interaction. Then, our multi-band reference system
follows from
\begin{equation}
\hat{H}_{\rm H}=\hat{H}_{\rm kin} + \hat{V}_{\rm loc}-\hat{V}_{\rm dc}\; ,
\label{eq:HubbardHamiltonian}
\end{equation}
where $\hat{V}_{\rm loc}$ describes local interactions between electrons
in Wannier orbitals on the same site~$\vecR$. 
The local single-particle operator $\hat{V}_{\rm dc}$ 
accounts for the double counting of their interactions
in the Hubbard term $\hat{V}_{\rm loc}$ and 
in the single-particle exchange-correlation energy $E_{\rm sp,xc}$.
We assume that $\hat{V}_{\rm loc}$ and $\hat{V}_{\rm dc}$ 
do not depend on the densities $n_{\sigma}(\vecr)$ explicitly.

For the local interaction we set
\begin{eqnarray}
\hat{V}_{\rm loc} &=& \sum_{\vecR} \hat{V}_{\rm loc}(\vecR) \; , \nonumber \\
\hat{V}_{\rm loc}(\vecR) 
&=&  
\frac{1}{2} \sum_{(c_1,\sigma_1),\ldots,(c_4,\sigma_4)}
U^{(c_1,\sigma_1), (c_2,\sigma_2)}_{(c_3,\sigma_3), (c_4,\sigma_4)}\,
\hat{c} _{\vecR,c_1,\sigma_1}^{\dagger} \hat{c} _{\vecR,c_2,\sigma_2}^{\dagger} 
\hat{c} _{\vecR,c_3,\sigma_3}^{\vphantom{\dagger}}
 \hat{c} _{\vecR,c_4,\sigma_4}^{\vphantom{\dagger}} 
\label{eq:VlocR}\; .
\end{eqnarray}
Note that only electrons in the small subset of correlated 
orbitals (index~$c$) experience the two-particle interaction $\hat{V}_{\rm loc}$:
When there are two electrons in the Wannier orbitals $\phi_{\vecR,c_3,\sigma_3}(\vecr)$
and $\phi_{\vecR,c_4,\sigma_4}(\vecr)$ centered around the lattice site~$\vecR$,
they are scattered into the orbitals
$\phi_{\vecR,c_1,\sigma_1}(\vecr)$
and $\phi_{\vecR,c_2,\sigma_2}(\vecr)$, 
centered around the same lattice site~$\vecR$; 
for the definition of basis states, see~\ref{app:basissets}.
Typically, we consider $c=3d$ for the transition metals and their compounds.

The interaction strengths are parameters of the theory. 
Later, we shall employ the spherical approximation so that $U^{\cdots}_{\cdots}$
for $d$-electrons
can be expressed in terms of three Racah parameters $A$, $B$, and $C$.
Fixing $C/B$ makes it
possible to introduce an effective Hubbard parameter~$U$ and an effective
Hund's-rule coupling~$J$, see Sect.~\ref{sec:resultsnickel} and \hbox{\ref{app:UJ}}.
%
%
Due to screening, the effective Hubbard interaction~$U$ 
is  smaller than its bare, atomic value. In general, 
$U$ and $J$ are chosen to obtain good agreement with experiment,
see Sect.~\ref{sec:resultsnickel}.
%
%

\subsubsection{Hubbard density functional.}

According to Levy's constrained search, we must find the minimum
of the functional
\begin{equation}
F_{\rm H}\left[\left\{ n_{\sigma}(\vecr)\right\},\left\{|\Psi^{(n)}\rangle\right\}\right]
= \langle \Psi^{(n)} | \hat{H}_{\rm H} | \Psi^{(n)} \rangle 
\label{eq:RitzHubbardDFT}
\end{equation}
in the subset of normalized states 
with given (physical) density $n_{\sigma}(\vecr)$, see eq.~(\ref{eq:electrondensity}).
The minimum of 
$F_{\rm H}\left[\left\{ n_{\sigma}(\vecr)\right\},\left\{|\Psi^{(n)}\rangle\right\}\right]$
over the states $|\Psi^{(n)}\rangle$
is the ground state $|\Psi^{(n)}_{{\rm H},0}\rangle$
of the Hamiltonian $\hat{H}_{\rm H}$ 
for fixed densities $n_{\sigma}(\vecr)$.
In analogy to Sect.~\ref{subsec:SPHamiltonian}, we 
define the Hubbard density functional
\begin{eqnarray}
D_{\rm H}\left[\left\{ n_{\sigma}(\vecr) \right\}\right]
&=& 
K_{\rm H}\left[\left\{ n_{\sigma}(\vecr) \right\}\right] 
+ U\left[\left\{ n_{\sigma}(\vecr) \right\}\right] 
+  V_{\rm Har}\left[\left\{ n_{\sigma}(\vecr) \right\}\right] \nonumber \\
&&+ V_{\rm loc}\left[\left\{ n_{\sigma}(\vecr) \right\}\right] -
V_{\rm dc}\left[\left\{ n_{\sigma}(\vecr) \right\}\right] 
+  E_{\rm H,xc}\left[\left\{ n_{\sigma}(\vecr) \right\}\right] \; ,
\end{eqnarray}
where
\begin{eqnarray}
K_{\rm H}\left[\left\{ n_{\sigma}(\vecr) \right\}\right] 
&=&\langle \Psi_{\rm H,0}^{(n)} | \hat{H}_{\rm kin}| \Psi_{\rm H,0}^{(n)} \rangle 
\nonumber \; , 
\\
V_{\rm loc/dc}\left[\left\{ n_{\sigma}(\vecr) \right\}\right]
&=& \langle \Psi_{\rm H,0}^{(n)} | \hat{V}_{\rm loc/dc}| \Psi_{\rm H,0}^{(n)} \rangle   \; ,
\label{eq:functionsHubbard}
\end{eqnarray}
and $E_{\rm H,xc}\left[\left\{ n_{\sigma}(\vecr) \right\}\right] $
is the exchange-correlation energy for $\hat{H}_{\rm H}$.
As in Sect.~\ref{subsec:SPHamiltonian},
the Hubbard density functional agrees with the 
exact density functional if we choose
\begin{eqnarray}
E_{\rm H,xc}\left[\left\{ n_{\sigma}(\vecr) \right\}\right] 
&=& K\left[\left\{ n_{\sigma}(\vecr) \right\}\right] 
-K_{\rm H}\left[\left\{ n_{\sigma}(\vecr) \right\}\right] 
\nonumber \\
&& 
+ E_{\rm xc}\left[\left\{ n_{\sigma}(\vecr) \right\}\right] 
- \left( 
V_{\rm loc}\left[\left\{ n_{\sigma}(\vecr) \right\}\right] -
V_{\rm dc}\left[\left\{ n_{\sigma}(\vecr) \right\}\right] 
\right) 
\label{eq:subtractExc} \; .
\end{eqnarray}
Then, the Hubbard approach provides the exact ground-state densities and
ground-state energy of our full many-particle Hamiltonian
(Hubbard--Hohenberg-Kohn theorem).
Of course, our derivation relies on the assumption of Hubbard $V$-representability
of the densities~$n_{\sigma}(\vecr)$.

\subsubsection{Hubbard single-particle potential.}

When we directly apply 
the Ritz principle, we have to minimize the energy functional
$E\equiv E\left[ \left\{ n_{\sigma}(\vecr) \right\},\left\{ |\Psi\rangle \right\}\right]$
\begin{equation}
E=\langle \Psi | \hat{H}_{\rm H} | \Psi \rangle
+ U\left[\left\{ n_{\sigma}(\vecr) \right\}\right] 
+  V_{\rm Har}\left[\left\{ n_{\sigma}(\vecr) \right\}\right]
+  E_{\rm H,xc}\left[\left\{ n_{\sigma}(\vecr) \right\}\right]  \;.
\label{eq:Hubbardenergy}
\end{equation}
We include the constraints eq.~(\ref{eq:electrondensity}) and the normalization
condition using the Lagrange parameters $\llambda_{\sigma}(\vecr)$ and~$E_0$ 
in the functional $G_{\rm H}\equiv G_{\rm H}\left[ \left\{ |\Psi\rangle \right\}, 
\left\{ n_{\sigma}(\vecr) \right\},\left\{\llambda_{\sigma}(\vecr)\right\}, E_0\right]$
\begin{eqnarray}
G_{\rm H}
&=& E\left[ \left\{ n_{\sigma}(\vecr) \right\},\left\{ |\Psi\rangle \right\}\right]
-E_0\left(\langle \Psi |  \Psi \rangle-1\right) 
\nonumber \\
&& 
-\sum_{\sigma}\int{\rm d}\vecr \llambda_{\sigma}(\vecr)
\left(
n_{\sigma}(\vecr) - \langle \Psi | 
\hat{\Psi}_{\sigma}^{\dagger}(\vecr)\hat{\Psi}_{\sigma}^{\vphantom{\dagger}}(\vecr)  
| \Psi \rangle
\right)
\; .
\label{eq:Gweaddress}
\end{eqnarray}
As in Sect.~\ref{subsubsec:Kohn-Sham}, see eqs.~(\ref{eq:lambdaissppotential})
and~(\ref{eq:defineVKSsss}),
the variation of 
$G_{\rm H}$
with respect to $n_{\sigma}(\vecr)$ gives the single-particle potential
\begin{eqnarray}
V_{\sigma}^{\rm H}(\vecr)
&\equiv& U(\vecr)+V_{\rm Har}(\vecr)+v_{{\rm H,xc},\sigma}(\vecr)
\nonumber \; ,\\
v_{{\rm H,xc},\sigma}(\vecr) &\equiv& 
\left.
\frac{\partial E_{\rm H,xc}\left[\left\{ n_{\sigma'}(\vecr') \right\}\right]}%
{\partial n_{\sigma}(\vecr)} 
\right|_{n_{\sigma}(\vecr)=n_{\sigma}^0(\vecr)}
\label{eq:getmeHubbard}
 \; .
\end{eqnarray}
The Hubbard-model approach is based on the idea that 
typical approximations for the exchange-correlation energy, 
e.g., the local-density approximation,
are suitable for the Hubbard model,
\begin{equation}
E_{\rm H,xc}\left[\left\{ n_{\sigma}(\vecr) \right\}\right] 
\approx E_{\rm LDA,xc}\left[\left\{ n_{\sigma}(\vecr) \right\}\right]\; .
\label{eq:EHxcisExcLDA}
\end{equation}
Indeed, as seen from eq.~(\ref{eq:subtractExc}),
in the Hubbard exchange-correlation energy $E_{\rm H,xc}$ 
the exchange-correlation
contributions in the exact $E_{\rm xc}$ are reduced by the Hubbard term
$V_{\rm loc}\left[\left\{ n_{\sigma}(\vecr) \right\}\right] -
V_{\rm dc}\left[\left\{ n_{\sigma}(\vecr) \right\}\right] $,
reflecting a more elaborate treatment of local correlations.

The minimization of~(\ref{eq:Hubbardenergy}) with respect to $| \Psi \rangle$
constitutes an unsolvable many-particle problem. The 
ground state $|\Psi_0\rangle$ is the 
solution of the many-particle Schr\"odinger equation with energy $E_0$,
\begin{equation}
\left(\hat{H}_0 + \hat{V}_{\rm loc}-\hat{V}_{\rm dc}\right) |\Psi_0 \rangle
= E_0 |\Psi_0 \rangle
\label{eq:HMgroundstateeq}
\end{equation}
with the single-particle Hamiltonian
\begin{equation}
\hat{H}_0 =\sum_{\sigma} 
\int {\rm d} \vecr 
\hat{\Psi}_{\sigma}^{\dagger}(\vecr)\left( 
-\frac{\Delta_{\vecr}}{2m}  + 
U(\vecr)+V_{\rm Har}(\vecr)+v_{{\rm H,xc},\sigma}(\vecr)
\right)
\hat{\Psi}_{\sigma}^{\vphantom{\dagger}}(\vecr)   \; .
\label{eq:defineh0}
\end{equation}
The Schr\"odinger equation~(\ref{eq:HMgroundstateeq})
can be used as starting point for further approximations,
for example, the Dynamical Mean-Field Theory (DMFT).
In the following we will address the functional in eq.~(\ref{eq:Hubbardenergy})
directly.

\subsection{Gutzwiller density functional}
\label{subsec:G-DFT}

In the widely used LDA+$U$ approach~\cite{Lichtenstein}, 
the functional in eq.~(\ref{eq:Hubbardenergy})
is evaluated and (approximately) minimized
by means of single-particle product wave functions.
However, this approach treats correlations only on a mean-field level.
In the more sophisticated Gutzwiller approach, 
we consider the functional in eq.~(\ref{eq:Hubbardenergy})
in the subset of Gutzwiller-correlated variational many-particle states.

\subsubsection{Gutzwiller variational ground state.}
\label{subsec:Gutzi}

In order to formulate the 
Gutzwiller variational ground state~\cite{Gutzwiller,buenemann2005},
we consider the local (atomic) states $|\Gamma\rangle_{\vecR}$
that are built from the correlated orbitals.
The local Hamiltonians take the form
\begin{equation}
\hat{V}_{\rm loc/dc}(\vecR) = \sum_{\Gamma,\Gamma'} E_{\Gamma,\Gamma'}^{\rm loc/dc}(\vecR)
|\Gamma \rangle_{\vecR} {}_{\vecR} \langle \Gamma' | 
= \sum_{\Gamma,\Gamma'} E_{\Gamma,\Gamma'}^{\rm loc/dc}(\vecR) 
\hat{m}_{\vecR;\Gamma,\Gamma'} 
\; ,
\label{eq:Vloclocal}
\end{equation}
where $|\Gamma\rangle_{\vecR}$ contains $|\Gamma_{\vecR}|$ electrons.
Here, we introduced
\begin{equation}
E_{\Gamma,\Gamma'}^{\rm loc/dc}(\vecR)=
{}_{\vecR}\langle \Gamma| \hat{V}_{\rm loc/dc}(\vecR)| \Gamma'\rangle_{\vecR}
\end{equation}
and the local many-particle operators
$\hat{m}_{\vecR;\Gamma,\Gamma'} 
= |\Gamma\rangle_{\vecR} {}_{\vecR} \langle \Gamma' |$.

The Gutzwiller correlator and the Gutzwiller variational states
are defined as
\begin{equation}
\hat{P}_{\rm G}= \prod_{\vecR} \sum_{\Gamma,\Gamma'}\lambda_{\Gamma,\Gamma'}(\vecR)
\hat{m}_{\vecR;\Gamma,\Gamma'}
\quad , \quad 
|\Psi_{\rm G}\rangle= \hat{P}_{\rm G} | \Phi \rangle \; .
\end{equation}
Here, $| \Phi \rangle$ is a single-particle product state,
and $\lambda_{\Gamma,\Gamma'}(\vecR)$
defines the matrix $\tilde{\lambda}(\vecR)$ of, in general, complex 
variational parameters.

\subsubsection{Gutzwiller functionals.}
We evaluate the energy functional~(\ref{eq:Hubbardenergy})
in the restricted subset of Gutzwiller variational states,
\begin{eqnarray}
E\left[\left\{n_{\sigma}(\vecr)\right\} , \left\{|\Psi_{\rm G}\rangle\right\} \right] 
&=& 
\sum_{\vecR,b,\vecR',b',\sigma} T_{(\vecR,b),(\vecR',b');\sigma}
\rho_{(\vecR',b'),(\vecR,b);\sigma}^{\rm G} +V_{\rm loc}^{\rm G}
-V_{\rm dc}^{\rm G} \nonumber 
\\
&& + U\left[\left\{ n_{\sigma}(\vecr) \right\}\right] 
+  V_{\rm Har}\left[\left\{ n_{\sigma}(\vecr) \right\}\right] 
+  E_{\rm H,xc}\left[\left\{ n_{\sigma}(\vecr) \right\}\right]  ,
\nonumber \\[6pt]
\hphantom{|\Psi_{\rm G}\rangle, n_{\sigma}(\vecr)}
V_{\rm loc/dc}^{\rm G} &=&
\sum_{\vecR} \sum_{\Gamma,\Gamma'} E_{\Gamma,\Gamma'}^{\rm loc/dc}(\vecR)
m_{\vecR;\Gamma,\Gamma'}^{\rm G} \; . \label{eq:Gutzwenergy}
\label{eq:GutzenergywithUVExc}
\label{eq:CDCenergy}
\end{eqnarray}
Note that we work with the orbital Wannier basis, see~\ref{app:basissets},
\begin{equation}
T_{(\vecR,b),(\vecR',b');\sigma} = \int {\rm d}\vecr
\phi_{\vecR,b,\sigma}^*(\vecr)
\left(-\frac{\Delta_{\vecr}}{2m} \right)
\phi_{\vecR',b',\sigma}(\vecr) \; .
\end{equation}
The elements of the Gutzwiller-correlated single-particle density matrix are
\begin{equation}
\rho_{(\vecR',b'),(\vecR,b);\sigma}^{\rm G}=
\frac{\langle \Psi_{\rm G} | 
\hat{c}_{\vecR,b,\sigma}^{\dagger} \hat{c}_{\vecR',b',\sigma}^{\vphantom{\dagger}}
| \Psi_{\rm G} \rangle}{
\langle \Psi_{\rm G} | \Psi_{\rm G} \rangle} 
= \frac{\langle \Phi | \hat{P}_{\rm G}^{\dagger}
\hat{c}_{\vecR,b,\sigma}^{\dagger} \hat{c}_{\vecR',b',\sigma}^{\vphantom{\dagger}}
\hat{P}_{\rm G}^{\vphantom{\dagger}}
| \Phi \rangle}%
{\langle \Phi | \hat{P}_{\rm G}^{\dagger}\hat{P}_{\rm G}^{\vphantom{\dagger}}
|\Phi \rangle} 
\; ,
\label{eq:correlatedrho}
\end{equation}
and the densities become
\begin{equation}
n_{\sigma}(\vecr)= \sum_{\vecR,b,\vecR',b'} \phi_{\vecR,b,\sigma}^*(\vecr)
\phi_{\vecR',b',\sigma}(\vecr)
\rho_{(\vecR',b'),(\vecR,b);\sigma}^{\rm G}
\; .
\label{eq:Gutzdensities}
\end{equation}
The expectation values for the atomic operators are given by
\begin{equation}
m_{\vecR;\Gamma,\Gamma'}^{\rm G}
= 
\frac{\langle \Psi_{\rm G} | \hat{m}_{\vecR;\Gamma,\Gamma'}
| \Psi_{\rm G} \rangle}{
\langle \Psi_{\rm G} | \Psi_{\rm G} \rangle} 
=
\frac{\langle \Phi | \hat{P}_{\rm G}^{\dagger}
\hat{m}_{\vecR;\Gamma,\Gamma'}
\hat{P}_{\rm G}^{\vphantom{\dagger}} | \Phi \rangle}%
{\langle \Phi | \hat{P}_{\rm G}^{\dagger}\hat{P}_{\rm G}^{\vphantom{\dagger}}
|\Phi \rangle}  \; .
\end{equation}
The diagrammatic evaluation of 
$\rho_{(\vecR',b'),(\vecR,b);\sigma}^{\rm G}$ and of $m_{\vecR;\Gamma,\Gamma'}^{\rm G}$
shows that
these quantities are functionals of the non-interacting single-particle density matrices
$\tilde{\rho}$, see eq.~(\ref{eq:rhoelements}),
and of the variational parameters $\lambda_{\Gamma,\Gamma'}(\vecR)$.
Moreover, it turns out that the local, non-interacting single-particle 
density matrix $\tilde{C}(\vecR)$ with the elements
\begin{equation}
C_{b,b';\sigma}(\vecR) \equiv \rho_{(\vecR,b),(\vecR,b');\sigma}
\label{eq:etasbrauchenwirhier}
\end{equation}
plays a prominent role in the Gutzwiller energy functional, in particular,
for infinite lattice coordination number.
Therefore, we may write
\begin{equation}
E\left[\left\{ n_{\sigma}(\vecr) \right\},\left\{|\Psi_{\rm G}\rangle\right\}\right]
\equiv 
E^{\rm G}
\left[ \tilde{\rho},\left\{\tilde{\lambda}(\vecR)\right\},
\left\{ n_{\sigma}(\vecr) \right\},
\left\{ \tilde{C}(\vecR) \right\}
\right] \; .
\label{eq:sixtysix}
\end{equation}
In the Lagrange functional we shall impose the relation~(\ref{eq:etasbrauchenwirhier})
with the help of the Hermitian
Lagrange parameter matrix $\tilde{\eta}$ 
with entries $\eta_{(\vecR,b),(\vecR,b');\sigma}$.
Lastly, for the analytical evaluation of eq.~(\ref{eq:sixtysix}) it is helpful
to impose a set of (real-valued) local constraints  ($l=1,2,\ldots,N_{\rm con}$)
\begin{equation}
g_{l,\vecR} \left[\tilde{\lambda}(\vecR),\tilde{C}(\vecR)\right] = 0  \; ,
\label{eq:constraints}
\end{equation}
which we implement with real 
Lagrange parameters $\Lambda_l(\vecR)$;
%
%
for explicit expressions, see eqs.~(\ref{5.5}) and~(\ref{5.5b}).
%
%

In the following, we abbreviate $i=(\vecR,b)$ and $j=(\vecR',b')$. 
Consequently, in analogy with Sect.~\ref{subsubsec:Kohn-Sham}, we 
address 
\begin{equation}
G_{\rm DFT}^{\rm G}\equiv 
G_{\rm DFT}^{\rm G}\left[
{\arraycolsep=1pt\begin{array}{@{}llll@{}}
\tilde{\rho},&\left\{n_{\sigma}(\vecr)\right\},&\left\{ \tilde{C}(\vecR) \right\},
&\left\{\tilde{\lambda}(\vecR)\right\}\\
\widetilde{\Omega},&\left\{ \llambda_{\sigma}(\vecr)\right\},
&\left\{\tilde{\eta}(\vecR) \right\},
&\left\{\Lambda_l(\vecR)\right\}
\end{array}}
\right]
\end{equation}
as our Lagrange functional,
\begin{eqnarray}
G_{\rm DFT}^{\rm G}
&=&
E^{\rm G}
\left[ \tilde{\rho},\left\{\tilde{\lambda}(\vecR)\right\},\left\{ n_{\sigma}(\vecr) \right\},
\left\{\tilde{C}(\vecR)\right\}\right]
-\sum_{l,m,\sigma}\Omega_{l,m;\sigma} 
\left( \tilde{\rho}\cdot \tilde{\rho}-\tilde{\rho}\right)_{m,l;\sigma}
 \nonumber \\
&&-\sum_{\sigma} \int{\rm d}\vecr \llambda_{\sigma}(\vecr) 
\biggl(n_{\sigma}(\vecr)- \sum_{i,j} 
\phi_{i,\sigma}^*(\vecr)\phi_{j,\sigma}(\vecr)
\rho_{j,i;\sigma}^{\rm G}\biggr)  
\label{eq:minimizewithlambda-Gutz}\\
&& +\sum_{l,\vecR}\Lambda_l(\vecR) 
g_{l,\vecR}
- \sum_{\vecR,b,b',\sigma}\eta_{b,b';\sigma}(\vecR)\left(C_{b',b;\sigma}(\vecR)-
\rho_{(\vecR,b'),(\vecR,b);\sigma} \right)\; ,
\nonumber 
\end{eqnarray}
cf.~eq.~(\ref{eq:minimizewithlambda}).
Here, we took the condition~(\ref{eq:Gutzdensities}) into account
using Lagrange parameters $\llambda_{\sigma}(\vecr)$ because 
the external potential, the Hartree term and the exchange-correlation potential
in eq.~(\ref{eq:GutzenergywithUVExc}) depend on the densities.

\subsubsection{Minimization of the Gutzwiller energy functional.}
\label{subsubsec:qpHamilt}

The functional~$G_{\rm DFT}^{\rm G}$ in eq.~(\ref{eq:minimizewithlambda-Gutz})
has to be minimized with respect to $n_{\sigma}(\vecr)$, $\tilde{C}(\vecR)$,
$\tilde{\lambda}(\vecR)$, and $\tilde{\rho}$.
The variation with respect to the Lagrange parameters 
$\kappa_{\sigma}(\vecr)$, $\tilde{\eta}(\vecR)$, $\Lambda_l(\vecR)$, 
and $\widetilde{\Omega}$
gives the constraints (\ref{eq:Gutzdensities}), 
(\ref{eq:etasbrauchenwirhier}), 
(\ref{eq:constraints}), and 
(\ref{eq:rhorho}), respectively.
\begin{enumerate}
\item 
As in the derivation of the exact Schr\"odinger equation~(\ref{eq:HMgroundstateeq}),
the variation of $G_{\rm DFT}^{\rm G}$ with respect to $n_{\sigma}(\vecr)$
generates the single-particle potential,
\begin{equation}
\llambda_{\sigma}(\vecr)=V_{\sigma}^{\rm H}(\vecr)\; ,
\end{equation}
see eqs.~(\ref{eq:defineVKSsss}) and~(\ref{eq:getmeHubbard}).
\item 
The minimization with respect to 
$\tilde{C}(\vecR)$ gives
\begin{eqnarray}
\eta_{b,b';\sigma}(\vecR) &=& \frac{\partial  E^{\rm G}}{\partial C_{b',b;\sigma}(\vecR)}
+ \sum_{l} \Lambda_l(\vecR) \frac{\partial  g_{l,\vecR}}{\partial C_{b',b;\sigma}(\vecR)}
\nonumber \\
&& +
\sum_{i,j,\sigma'} \int{\rm d}\vecr V_{\sigma'}^{\rm H}(\vecr) 
\phi_{i,\sigma'}^*(\vecr)\phi_{j,\sigma'}(\vecr)
\frac{\partial \rho_{j,i;\sigma'}^{\rm G}}{\partial C_{b',b;\sigma}(\vecR)}
\; .
\label{eq:etaallgemeien}
\end{eqnarray}
\item 
The minimization with respect to the Gutzwiller correlation parameters 
$\tilde{\lambda}(\vecR)$ results in
\begin{eqnarray}
0 &=& 
\frac{\partial E^{\rm G}}{\partial \lambda_{\Gamma,\Gamma'}(\vecR)}
+\sum_{l,m,\sigma}\int {\rm d}\vecr V_{\sigma}^{\rm H}(\vecr)
\phi_{l,\sigma}^*(\vecr)\phi_{m,\sigma}(\vecr)
  \frac{\partial \rho_{m,l,\sigma}^{\rm G}}{\partial \lambda_{\Gamma,\Gamma'}(\vecR)}
\nonumber \\
&& +\sum_{l} \Lambda_l(\vecR) 
\frac{\partial g_{l,\vecR}}{\partial \lambda_{\Gamma,\Gamma'}(\vecR)}
\\
&=& 
\sum_{l,m,\sigma} \!h_{l,m;\sigma}^0
  \frac{\partial \rho_{m,l,\sigma}^{\rm G}}{\partial \lambda_{\Gamma,\Gamma'}(\vecR)}
+ 
\frac{\partial \left(V_{\rm loc}^{\rm G}-V_{\rm dc}^{\rm G}\right)}%
{\partial \lambda_{\Gamma,\Gamma'}(\vecR)}
+\!\sum_{l} \Lambda_l(\vecR) 
\frac{\partial g_{l,\vecR}}{\partial \lambda_{\Gamma,\Gamma'}(\vecR)}
\nonumber \,,
\\
h_{l,m;\sigma}^0&\equiv& \int {\rm d}\vecr 
\phi_{l,\sigma}^*(\vecr) 
\Bigl(-\frac{\Delta_{\vecr}}{2m}  + 
U(\vecr)+V_{\rm Har}(\vecr)+v_{{\rm H,xc},\sigma}(\vecr)
\Bigr)
\phi_{m,\sigma}(\vecr)
\label{eq:h0defhere}
\end{eqnarray}
for all $\lambda_{\Gamma,\Gamma'}(\vecR)$.
Note that, in the case of complex Gutzwiller parameters,
we also have to minimize with respect to 
$(\lambda_{\Gamma,\Gamma'}(\vecR))^*$.
Using these equations we may calculate the Lagrange parameters $\Lambda_l(\vecR)$
that are needed in eq.~(\ref{eq:etaallgemeien}).
\item 
The minimization of $G_{\rm DFT}^{\rm G}$ with respect to $\tilde{\rho}$
generates the Landau--Gutzwiller quasi-particle Hamiltonian, 
see~\ref{appbueni},
\begin{equation}
\hat{H}_{\rm qp}^{\rm G}=\sum_{i,j,\sigma} h_{i,j;\sigma}^{\rm G} 
\hat{c}_{i,\sigma}^{\dagger}\hat{c}_{j,\sigma}^{\vphantom{\dagger}}
\label{eq:qpGutzhamiltonian}
\end{equation}
with the entries
\begin{eqnarray}
h_{i,j;\sigma}^{\rm G} &=& 
\frac{\partial E^{\rm G}}{\partial \rho_{j,i;\sigma}}
+\sum_{l,m,\sigma'}\int {\rm d}\vecr V_{\sigma'}^{\rm H}(\vecr)
\phi_{l,\sigma'}^*(\vecr)\phi_{m,\sigma'}(\vecr)
  \frac{\partial \rho_{m,l,\sigma'}^{\rm G}}{\partial \rho_{j,i;\sigma}}\nonumber
\\
&& + \sum_{\vecR,b,b',\sigma'} 
\eta_{b,b';\sigma'}(\vecR) 
 \frac{\partial \rho_{(\vecR,b'),(\vecR,b);\sigma'}}{\partial \rho_{j,i;\sigma}}\nonumber
\nonumber \\
&=& 
\sum_{l,m,\sigma'} h_{l,m;\sigma'}^0
  \frac{\partial \rho_{m,l,\sigma'}^{\rm G}}{\partial \rho_{j,i;\sigma}}\nonumber
+ \frac{\partial \left(V_{\rm loc}^{\rm G}-V_{\rm dc}^{\rm G}\right)}{\partial \rho_{j,i;\sigma}}
\\
&& 
+ \sum_{\vecR,b,b'} \delta_{j,(\vecR,b')}\delta_{i,(\vecR,b)}\eta_{b,b';\sigma}(\vecR) 
\;,  \label{eq:getGutzwillerTij} 
\end{eqnarray}
where we used eqs.~(\ref{eq:Gutzwenergy}) and~(\ref{eq:h0defhere}).

The single-particle state $|\Phi\rangle$ is the ground state of the
Hamiltonian~(\ref{eq:qpGutzhamiltonian}) from which the single-particle
density matrix $\tilde{\rho}$ follows.
\end{enumerate}
The minimization problem outlined in steps~(i)--(iv) 
requires the evaluation of the energy $E^{\rm G}$ in eq.~(\ref{eq:Gutzwenergy}).
In particular, the correlated single-particle density matrix $\tilde{\rho}^{\rm G}$,
eq.~(\ref{eq:correlatedrho}), must be determined.

All equations derived in this section are completely general. They can, at least
in principle, be evaluated by means of a diagrammatic expansion 
method~\cite{GebhardPRB90,GebhardBuenemannSchickling2012,Kacmarcyk}.
The leading order of the expansion corresponds to an approximation-free
evaluation of expectation values for Gutzwiller wave functions
in the limit of high lattice coordination number. 
This limit will be studied in the rest of this work.

\subsection{Gutzwiller density functional for infinite lattice coordination number}
\label{subsec:inftysimplify}

For $Z\to \infty$, 
the Gutzwiller-correlated single-particle 
density matrix and the Gutzwiller probabilities for the local occupancies can be
calculated explicitly without further approximations.
%
%
In this section we make no symmetry assumptions
(translational invariance, crystal symmetries). Note, however, that the equations 
do not cover the case of spin-orbit coupling.
%
%

\subsubsection{Local constraints.}

As shown in Refs.~\cite{buenemann2005,buenemann1998}
it is convenient for the evaluation of Gutzwiller wave functions 
to impose the following (local) constraints
\begin{equation}
\label{5.5}
\sum_{\Gamma,\Gamma_1,\Gamma_2}
\lambda_{\Gamma,\Gamma_1}^*(\vecR)\lambda_{\Gamma,\Gamma_2}(\vecR)
\langle
\hat{m}_{\vecR;\Gamma_1,\Gamma_2} \rangle_{\Phi} 
=1
\end{equation}
and
\begin{equation}
\label{5.5b}
\sum_{\Gamma,\Gamma_1,\Gamma_2}
\lambda_{\Gamma,\Gamma_1}^*(\vecR)\lambda_{\Gamma,\Gamma_2}(\vecR)
\langle  \hat{m}_{\vecR; \Gamma_1,\Gamma_2}
\hat{c}_{\vecR,b,\sigma}^{\dagger}
\hat{c}_{\vecR,b',\sigma}^{\vphantom{\dagger}}
\rangle_{\Phi} 
=
\langle \hat{c}_{\vecR,b,\sigma}^{\dagger}
\hat{c}_{\vecR,b',\sigma}^{\vphantom{\dagger}}
\rangle_{\Phi}
 \;,\end{equation}
where  we abbreviated $\langle \hat{A} \rangle_{\Phi}\equiv
\langle \Phi |\hat{A}| \Phi \rangle$.
Note that, for complex constraints, the index~$l$ in~(\ref{eq:constraints})
labels real and imaginary parts separately.

\subsubsection{Atomic occupancies.}
In the limit of infinite lattice coordination number,
the interaction and double-counting energy can be expressed solely
in terms of the local variational parameters $\tilde{\lambda}(\vecR)$
and the local density matrix $\tilde{C}(\vecR)$ of the 
correlated bands in $|\Phi\rangle$,
\begin{equation}
V_{\rm loc/dc}^{\rm G}
 =\sum_{\vecR} \sum_{\Gamma_1,\ldots,\Gamma_4} 
\lambda_{\Gamma_2,\Gamma_1}^*(\vecR) 
E_{\Gamma_2,\Gamma_3}^{\rm loc/dc}(\vecR)
\lambda_{\Gamma_3,\Gamma_4}(\vecR)  
\langle \hat{m}_{\vecR;\Gamma_1,\Gamma_4} \rangle_{\Phi} 
\; .
\label{eq:localenergy}
\end{equation}
The remaining expectation values 
$\langle \hat{m}_{\vecR;\Gamma_1,\Gamma_4} \rangle_{\Phi} $
are evaluated using Wick's theorem.
Explicit expressions are given in Refs.~\cite{buenemann2005,buenemann2012}.

\subsubsection{Correlated single-particle density matrix.}

The local part of the correlated single-particle density matrix
is given by
\begin{eqnarray}
\rho_{(\vecR,b'),( \vecR,b);\sigma}^{\rm G}
&=&\sum_{\Gamma_1,\ldots,\Gamma_4}
\lambda_{\Gamma_2,\Gamma_1}^*(\vecR)\lambda_{\Gamma_3,\Gamma_4}(\vecR)
\langle 
\hat{m}_{\vecR;\Gamma_1,\Gamma_2} 
\hat{c}_{\vecR,b,\sigma}^{\dagger}
\hat{c}_{\vecR,b',\sigma}^{\vphantom{\dagger}}
\hat{m}_{\vecR;\Gamma_3,\Gamma_4} 
\rangle_{\Phi}  \nonumber \\
&\equiv& C^{\rm G}_{b',b;\sigma}(\vecR) \; .
\label{eq:correlatedlocaldensity}
\end{eqnarray}
It can be evaluated using Wick's theorem. 
As can be seen from eq.~(\ref{eq:correlatedlocaldensity}),
it is a function of the variational parameters $\lambda_{\Gamma,\Gamma'}(\vecR)$
and of the local non-interacting single-particle density matrix $\tilde{C}(\vecR)$.

For $\vecR\neq \vecR'$, we have for the correlated single-particle density
matrix 
\begin{equation}
\rho_{(\vecR',b'),(\vecR,b);\sigma}^{\rm G}
= \sum_{a,a'}q_{b,\sigma}^{a,\sigma}(\vecR)
\left( q_{b',\sigma}^{a',\sigma}(\vecR')\right)^*  
\rho_{(\vecR',a'),(\vecR,a);\sigma}
\label{eq:qfactorsdef}
\end{equation}
with the orbital-dependent renormalization factors $q_{b,\sigma}^{a,\sigma}(\vecR)$
for the electron transfer between different sites.
Explicit expressions in terms of the variational parameters 
$\tilde{\lambda}(\vecR)$
and of the local non-interacting single-particle density matrix $\tilde{C}(\vecR)$
are given in Refs.~\cite{buenemann2005,buenemann2012}.

\section{Implementation for translational invariant systems}
\label{sec:transinvsystems}

For a system that is invariant under translation by a lattice vector
and contains only one atom per unit cell,
all local quantities become independent of the site index,
e.g., $\lambda_{\Gamma,\Gamma'}(\vecR)\equiv \lambda_{\Gamma,\Gamma'}$
for the Gutzwiller variational parameters. Since $\veck$ from the first Brillouin zone
is a good quantum number, we work with the (orbital) Bloch basis, 
see~\ref{app:basissets}.
%
%
It is straightforward to generalize the equations in Sect.~4 to the case of
more than one atom per unit cell. One simply has to add one more index that labels
the atoms in the unit cell. 
%
%

As shown in Sect.~\ref{subsubsec:qpHamilt},
the minimization of the Gutzwiller energy functional requires two major steps,
namely, the variation with respect to the Gutzwiller parameters $\tilde{\lambda}$
and the variation with respect to the single-particle density matrix
$\tilde{\rho}$ that characterizes the single-particle product state $|\Phi\rangle$.

\subsection{Gutzwiller--Kohn-Sham Hamiltonian}

The minimization of the energy functional with 
respect to the single-particle density matrix
leads to the Gutzwiller--Kohn-Sham Hamiltonian.
In the orbital Bloch basis $\phi_{\veck,b,\sigma}(\vecr)$,
see \ref{app:basissets},  
the corresponding quasi-particle Hamiltonian reads
\begin{equation}
\hat{H}_{\rm qp}^{\rm G}=\sum_{\veck,b,b',\sigma} 
h_{b,b';\sigma}^{\rm G}(\veck) 
\hat{c}_{\veck,b,\sigma}^{\dagger}\hat{c}_{\veck,b',\sigma}^{\vphantom{\dagger}}
\; ,
\label{eq:Hqpagain}
\end{equation}
see eq.~(\ref{eq:qpGutzhamiltonian}). In this section, we shall explain 
 how this Hamiltonian can be calculated. Note, however, that 
 the actual numerical implementation within {\sc QuantumEspresso} is done 
in first quantization and uses plane-waves. We therefore derive  
 the plane-wave representation of the Gutzwiller--Kohn-Sham equations 
in~\ref{subsubsec:pw-Gutzwiller}.

\subsubsection{Derivation of matrix elements.}
When we apply the general expressions~(\ref{eq:getGutzwillerTij}),
the matrix elements
of the quasi-particle Hamiltonian are obtained as
\begin{equation}
h_{b,b';\sigma}^{\rm G}(\veck) =
\eta_{b,b';\sigma}+ 
\sum_{a,a',\sigma'} h_{a,a';\sigma'}^0(\veck)
\frac{
\partial \rho_{a',a;\sigma'}^{\rm G}(\veck)
}{
\partial \rho_{b',b;\sigma}(\veck)
} \; ,
\label{eq:showetaimportance}
\end{equation}
where we have from eq.~(\ref{eq:h0defhere})
\begin{eqnarray}
h_{a,a';\sigma}^0(\veck) &=&
\int {\rm d}\vecr
\phi_{\veck,a,\sigma}^*(\vecr)
\left(-\frac{\Delta_{\vecr}}{2m} +
U(\vecr)+V_{\rm Har}(\vecr)+v_{{\rm H,xc},\sigma}(\vecr)
\right)
\phi_{\veck,a',\sigma}(\vecr) 
\, .
\nonumber \\
\end{eqnarray}
Moreover,
\begin{equation}
\rho_{b',b;\sigma}(\veck)
=\langle \Phi | 
 \hat{c}_{\veck,b,\sigma}^{\dagger} \hat{c}_{\veck,b',\sigma}^{\vphantom{\dagger}} 
| \Phi\rangle
\label{eq:defnbbprime}
\end{equation}
are the entries of the single-particle
density matrix in the orbital Bloch basis, and $\rho_{b',b;\sigma}^{\rm G}(\veck)$
denotes the corresponding quantities in the Gutzwiller-correlated state.
In the limit of infinite lattice coordination number, we may express
$V_{\rm loc/dc}^{\rm G}$ in eq.~(\ref{eq:getGutzwillerTij}) as a function of
the Gutzwiller variational parameters $\lambda_{\Gamma,\Gamma'}$
and of the local density matrix~$\tilde{C}$. Therefore, 
$V_{\rm loc/dc}^{\rm G}$ are formally 
independent of the single-particle density matrix $\tilde{\rho}$
so that they do not contribute
to $h_{b,b';\sigma}^{\rm G}(\veck) $.
Equation~(\ref{eq:showetaimportance}) shows that, 
apart from an overall shift of the orbitals through $\eta_{b,b';\sigma}$,
we can still work with the matrix elements
$h_{a,a';\sigma'}^0(\veck)$ of the single-particle operator $\hat{H}_0 $ that
enters the many-particle Schr\"odinger equation~(\ref{eq:HMgroundstateeq}).

In the orbital Bloch basis, eqs.~(\ref{eq:correlatedlocaldensity})
and~(\ref{eq:qfactorsdef}) take the form 
\begin{equation}
\rho_{b',b;\sigma}^{\rm G}(\veck)=
\sum_{a,a'}q_{b,\sigma}^{a,\sigma}\left(q_{b',\sigma}^{a',\sigma}\right)^*
\left(\rho_{a',a;\sigma}(\veck)
-  C_{a',a;\sigma} \right)+ 
C^{\rm G}_{b',b;\sigma}
\; .
\nonumber 
\label{eq:defnbbprimeGutzwiller}
\end{equation}
When we insert eq.~(\ref{eq:defnbbprimeGutzwiller})
into eq.~(\ref{eq:showetaimportance})
we thus find
for the entries of the Gutzwiller quasi-particle Hamiltonian
\begin{equation}
h_{b,b';\sigma}^{\rm G}(\veck) = \eta_{b,b';\sigma}
+ \sum_{a,a'}q_{a,\sigma}^{b,\sigma}\left(q_{a',\sigma}^{b',\sigma}\right)^*
h_{a,a';\sigma}^0(\veck)  \; . 
\label{eq:TbbprimeG}
\end{equation}
Recall that the local single-particle densities $\tilde{C}$
are treated as independent parameters. 
Since the $q$-factors and the correlated
local single-particle density matrix $C_{b,b';\sigma}^{\rm G}
=\rho_{(\vecR,b),( \vecR,b');\sigma}^{\rm G}$
are solely functions of the variational
parameters $\tilde{\lambda}$ and of $\tilde{C}$,
they are treated as constants when we take the derivative
of $\rho_{b',b;\sigma}^{\rm G}(\veck)$
with respect to $\rho_{b',b;\sigma}(\veck)$
in eq.~(\ref{eq:defnbbprimeGutzwiller}).

\subsubsection{Diagonalization of the quasi-particle Hamiltonian.}
The unitary matrix
$\tilde{F}_{\sigma}^{\rm G}(\veck)$
diagonalizes the Gutzwiller matrix $\tilde{h}_{\sigma}^{\rm G}(\veck)$,
\begin{equation}
\sum_{b,b'} (F_{b,{\rm n},\sigma}^{\rm G}(\veck) )^*
h_{b,b';\sigma}^{\rm G}(\veck) 
F_{b',{\rm m},\sigma}^{\rm G}(\veck) =
\delta_{\rm n,m}\epsilon_{{\rm n},\sigma}^{\rm G}(\veck)\; ,
\label{eq:qpdispersion}
\end{equation}
which provides the quasi-particle dispersion
$\epsilon_{{\rm n},\sigma}^{\rm G}(\veck)$.
We introduce the quasi-particle band operators
\begin{equation}
\hat{g}_{\veck,{\rm n},\sigma}^{\dagger} 
= \sum_b F_{b,{\rm n},\sigma}^{\rm G}(\veck) \hat{c}_{\veck,b,\sigma}^{\dagger}
\quad, \quad  
\hat{g}_{\veck,{\rm n},\sigma}^{\vphantom{\dagger}} 
= \sum_b (F_{b,{\rm n},\sigma}^{\rm G}(\veck))^* \hat{c}_{\veck,b,\sigma}^{\vphantom{\dagger}}
\; ,
\end{equation}
in which the Gutzwiller quasi-particle Hamiltonian becomes diagonal,
\begin{equation}
\hat{H}_{\rm qp}^{\rm G}
= \sum_{\veck,{\rm n},\sigma} \epsilon_{{\rm n},\sigma}^{\rm G}(\veck)
\hat{g}_{\veck,{\rm n},\sigma}^{\dagger}\hat{g}_{\veck,{\rm n},\sigma}^{\vphantom{\dagger}} \; .
\end{equation}
In order to minimize the Gutzwiller density functional $G_{\rm DFT}^{\rm G}$, we
must work with the ground state of 
the quasi-particle Hamiltonian $\hat{H}_{\rm qp}^{\rm G}$,
\begin{equation}
|\Phi_0\rangle = \prod_{\veck,{\rm n},\sigma}{}^{{}^{\prime}}
\hat{g}_{\veck,{\rm n},\sigma}^{\dagger}|\hbox{vac}\rangle \; ,
\end{equation}
where the $N$ levels lowest in energy are occupied as indicated by
the prime at the product, 
$\epsilon_{{\rm n},\sigma}^{\rm G}(\veck)\leq E_{{\rm F},\sigma}^{\rm G}$.
Using eq.~(\ref{eq:defnbbprime}) we find
\begin{eqnarray}
\rho_{b',b;\sigma}^{\rm opt}(\veck)&=& 
\langle \hat{c}_{\veck,b,\sigma}^{\dagger}\hat{c}_{\veck,b',\sigma}^{\vphantom{\dagger}}
\rangle_{\Phi_0} 
\nonumber \\
&=& \sum_{\rm n} 
(F_{b,{\rm n},\sigma}^{\rm G}(\veck))^* F_{b',{\rm n},\sigma}^{\rm G}(\veck) 
\langle \hat{g}_{\veck,{\rm n},\sigma}^{\dagger}\hat{g}_{\veck,{\rm n},\sigma}^{\vphantom{\dagger}}
\rangle_{\Phi_0}  \label{eq:qpdensities}\\
 &=& 
\sum_{\rm n}  f_{\veck,{\rm n},\sigma}^{\rm G} 
(F_{b,{\rm n},\sigma}^{\rm G}(\veck))^* F_{b',{\rm n},\sigma}^{\rm G}(\veck)  \; ,
\nonumber 
\end{eqnarray}
where the quasi-particle occupancies
\begin{equation}
f_{\veck,{\rm n},\sigma}^{\rm G} =
\langle \hat{g}_{\veck,{\rm n},\sigma}^{\dagger}
\hat{g}_{\veck,{\rm n},\sigma}^{\vphantom{\dagger}}
\rangle_{\Phi_0}  =
\Theta\left(E_{\rm F,\sigma}^{\rm G}-\epsilon_{{\rm n},\sigma}^{\rm G}(\veck)\right)
\end{equation}
are unity for occupied quasi-particle levels up to the 
Fermi energy~$E_{\rm F,\sigma}^{\rm G}$, and zero otherwise.
The particle densities follow from eq.~(\ref{eq:qpdensities}),
\begin{equation}
n_{\sigma}(\vecr) = \sum_{\veck,b,b'} \phi_{\veck,b,\sigma}^*(\vecr)
\phi_{\veck,b',\sigma}(\vecr) \rho_{b',b;\sigma}^{\rm G}(\veck)  \; ,
\label{eq:getdensities}
\end{equation}
where $\rho_{b',b;\sigma}^{\rm G}(\veck)$ 
is given by eq.~(\ref{eq:defnbbprimeGutzwiller}).
As in DFT, the particle densities must be calculated self-consistently.

\subsubsection{Band-shift parameters.}
In order to determine the band-shift parameters $\eta_{b,b';\sigma}$, we
must evaluate eq.~(\ref{eq:etaallgemeien}) using the Gutzwiller energy functional
in the limit of infinite lattice coordination number.
We define the kinetic energy of the Gutzwiller quasi-particles as
\begin{equation}
E_{\rm kin}^{\rm G}
=\sum_{\veck,{\rm n},\sigma} f_{\veck,{\rm n},\sigma}^{\rm G}
\epsilon_{{\rm n},\sigma}^{\rm G}(\veck)  
= \sum_{\veck,b,b',\sigma} h_{b,b';\sigma}^{\rm G}(\veck)
\rho_{b',b;\sigma}(\veck) 
\; .
\end{equation}
It is easy to show that
\begin{eqnarray}
\sum_{\veck,b,b',\sigma} h_{b,b';\sigma}^0(\veck) \rho_{b',b;\sigma}^{\rm G}(\veck)
&=& E_{\rm kin}^{\rm G}-
\sum_{\veck,b,b',\sigma} \eta_{b,b';\sigma}\rho_{b',b;\sigma}(\veck)
\nonumber \\
&& + L \sum_{b,b',\sigma} h_{b,b';\sigma}^0
\left(  C^{\rm G}_{b',b;\sigma}- C_{b',b;\sigma} \right) \; , \nonumber \\
\hphantom{h_{b,b';\sigma}^0(\veck) \rho_{b',b;\sigma}^{\rm G}(\veck)}
h_{b,b';\sigma}^0
&=& \frac{1}{L} \sum_{\veck} h_{b,b';\sigma}^0(\veck) \; .
\end{eqnarray}
Then, eq.~(\ref{eq:etaallgemeien}) gives
the effective local hybridizations $\eta_{b,b';\sigma}$
\begin{eqnarray}
L\eta_{b,b';\sigma}&=& 
 \frac{\partial}{\partial C_{b',b;\sigma}}
\biggl(
L\sum_l \Lambda_l g_l
+  V_{\rm loc}^{\rm G}- V_{\rm dc}^{\rm G}
+ E_{\rm kin}^{\rm G}
\nonumber \\
&&
\hphantom{ \frac{\partial}{\partial C_{b',b;\sigma}}\biggl(}
+L \sum_{a,a',\sigma'} \Bigl( C^{\rm G}_{a',a;\sigma}- C_{a',a;\sigma}
 \Bigr)h_{a,a';\sigma'}^0\biggr)
\; .
\label{eq:neededurgently}
\end{eqnarray}
Note that the term in the second line in eq.~(\ref{eq:neededurgently})
often vanishes due to symmetry, e.g., in nickel,
because $C^{\rm G}_{b,b';\sigma}=C_{b,b';\sigma}$.

\subsection{Minimization with respect to the Gutzwiller parameters}

In the (`inner') minimization with respect to the Gutzwiller parameters 
$\lambda_{\Gamma,\Gamma'}$  
(which are now independent of $\vecR$) we assume that the single-particle 
state $|\Phi \rangle$ is fixed.  
Then we have to minimize the function
\begin{eqnarray}\nonumber
&&E^{\rm inner}(\tilde{\lambda},\{\Lambda_l  \})\equiv
\sum_{\Gamma_1,\ldots,\Gamma_4}
\lambda_{\Gamma_2,\Gamma_1}^* 
(E_{\Gamma_2,\Gamma_3}^{\rm loc}-E_{\Gamma_2,\Gamma_3}^{\rm dc})
\lambda_{\Gamma_3,\Gamma_4}
\langle \hat{m}_{\Gamma_1,\Gamma_4} \rangle_{\Phi} \\\nonumber
&&+\sum_{\sigma} 
\sum_{c_1,c_2}\left[ \sum_{c_3,c_4}
q_{c_1,\sigma}^{c_2,\sigma}(\tilde{\lambda})
\left( q_{c_3,\sigma}^{c_4,\sigma}(\tilde{\lambda})\right)^*  I^{\sigma}_{c_1,c_3,c_2,c_4}
+\Big(q_{c_2,\sigma}^{c_1,\sigma}(\tilde{\lambda}) K^{\sigma}_{c_1,c_2}+{\rm c.c.}\Big)
\right]\\\label{dfhk}
&&+\sum_{\sigma}\sum_{b,b'}
h^{0}_{b,b';\sigma}C^{G}_{b',b;\sigma}(\tilde{\lambda})
+\sum_l\Lambda_lg_l(\tilde{\lambda}) \; ,
\end{eqnarray}
where we introduced
\begin{eqnarray}\label{ki1}
I^{\sigma}_{c_1,c_3,c_2,c_4}&\equiv&\frac{1}{L}
\sum_{\veck} h^{0}_{c_1,c_3;\sigma}(\veck)
(\rho_{c_4,c_2;\sigma}(\veck)-C_{c_4,c_2;\sigma}) \;, \\\label{ki2}
K^{\sigma}_{c,c'}&\equiv&\frac{1}{L}\sum_{\veck}\sum_{\bar{c},\bar{c}'}
h^{0}_{c,\bar{c};\sigma}(\veck)(\rho_{\bar{c}',c';\sigma}(\veck)-C_{\bar{c}',c';\sigma})\;.
\end{eqnarray}
Here, the indices $c$ and $\bar{c}$ denote correlated and non-correlated orbitals, 
respectively.  Note that the sum over $b,b'$ in the last line of eq.~(\ref{dfhk})
only contributes in the minimization 
if (at least) one of these two indices belongs 
to a correlated orbital. 
An efficient algorithm for the minimization of~(\ref{dfhk})
has been introduced in Ref.~\cite{buenemann2012}. This 
minimization also gives us the Lagrange parameters $\Lambda_l$ that enter the 
outer minimization in eq.~(\ref{eq:neededurgently}).

\section{Results for ferromagnetic nickel}
\label{sec:resultsnickel}

\subsection{Local Hamiltonian and double-counting corrections}
\label{ert}

For a  Gutzwiller DFT calculation we need to specify the Coulomb parameters in
the local Hamiltonian~(\ref{eq:VlocR}) and the form of the double-counting 
operator in~(\ref{eq:HubbardHamiltonian}). 

\subsubsection{Cubic symmetry and spherical approximation.}
In many theoretical studies one uses
a Hamiltonian with only density-density interactions,
\begin{eqnarray}
\label{app3.5}
\hat{V}^{\rm dens}_{\rm loc}&=&
\sum_{c,\sigma}U(c,c)
\hat{n}_{c,\sigma}\hat{n}_{c,\bar{\sigma}}
+\sum_{c(\neq)c'}\sum_{\sigma,\sigma'}
\widetilde{U}_{\sigma,\sigma'}(c,c')
\hat{n}_{c,\sigma}\hat{n}_{c',\sigma'} \, .
\end{eqnarray}
Here, we introduced $\bar{\uparrow}=\downarrow$ 
($\bar{\downarrow}=\uparrow$) and
 $\widetilde{U}_{\sigma,\sigma'}(c,c')= U(c,c')-\delta_{\sigma,\sigma'}J(c,c')$,
where $U(c,c')$ and  $J(c,c')$ are the local Hubbard and Hund's-rule exchange 
interactions. An additional and quite common approximation is the use
of orbital-independent Coulomb parameters,
\begin{equation}\label{jo1}
 U(c,c)\equiv U\; , \quad \hbox{and}\quad 
 U(c,c')\equiv  U',\; J(c,c')\equiv J \quad \hbox{for $c\neq c'.$}
\end{equation}
For a system of five correlated 3$d$ orbitals in a cubic environment as in nickel, 
however, the Hamiltonian~(\ref{app3.5})
is incomplete~\cite{Sugano1970}. The full Hamiltonian reads 
\begin{equation}
\hat{V}^{\rm full}_{\rm loc}
=\hat{V}^{\rm dens}_{\rm loc}+\hat{V}^{\rm n.dens.}_{\rm loc} \; ,
\end{equation}
 where
\begin{eqnarray}
\hat{V}^{\rm n.dens.}_{\rm loc}&=&\!
\sum_{c(\neq)c'}\!
J(c,c')\Bigl(\hat{c}^{\dagger}_{c,\uparrow}\hat{c}^{\dagger}_{c,\downarrow}
\hat{c}_{c',\downarrow}\hat{c}_{c',\uparrow}+ {\rm h.c.}\Bigr) 
+\!\!\sum_{c(\neq)c';\sigma}\!J(c,c')\hat{c}^{\dagger}_{c,\sigma}
\hat{c}^{\dagger}_{c',\bar{\sigma}}
\hat{c}_{c,\bar{\sigma}}\hat{c}_{c',\sigma}\nonumber\\
&&+\bigg[\sum_{t; \sigma,\sigma'}
(T(t)-\delta_{\sigma,\sigma'}A(t))
\hat{n}_{t,\sigma}\hat{c}^{\dagger}_{u,\sigma'}\hat{c}_{v,\sigma'}\nonumber\\
&&\hphantom{+\bigg[}
+\sum_{t,\sigma}A(t)
\left(
\hat{c}^{\dagger}_{t,\sigma}\hat{c}^{\dagger}_{t,\bar{\sigma}}
\hat{c}_{u,\bar{\sigma}}\hat{c}_{v,\sigma}+
\hat{c}^{\dagger}_{t,\sigma}\hat{c}^{\dagger}_{u,\bar{\sigma}}
\hat{c}_{t,\bar{\sigma}}\hat{c}_{v,\sigma}
\right)\nonumber \\
&&\hphantom{+\bigg[}
+\sum_{ t(\neq)t'(\neq)t^{\prime \prime}}
\sum_{e,\sigma,\sigma'}
S(t,t';t^{\prime \prime},e)
\hat{c}^{\dagger}_{t,\sigma}\hat{c}^{\dagger}_{t',\sigma'}
\hat{c}_{t^{\prime \prime},\sigma'}\hat{c}_{e,\sigma}+{\rm h.c.}\bigg ]\,.
\label{h255}
\end{eqnarray}
Here, $t=\zeta,\eta,\xi$ and $e=u,v$ are indices for the  
three $t_{2g}$ orbitals with symmetries $\zeta=xy$, $\eta=xz$, and $\xi=yz$,
and the two $e_g$ orbitals with symmetries
$u=3z^2-r^2$ and $v=x^2-y^2$, respectively. 
The parameters $A(t)$, $T(t)$, $S(t,t';t^{\prime \prime},e)$ 
in~eq.~(\ref{h255}) are of the same order of magnitude 
as the exchange interactions $J(c,c')$ and, hence, 
there is no a-priori reason to neglect $V_{\rm loc}^{\rm n.dens.}$.
Of all the parameters $U(c,c')$, $J(c,c')$,
$A(t)$, $T(t)$, $S(t,t';t^{\prime \prime},e)$ 
only ten are independent in cubic symmetry, see \ref{app:UJ}.

When we assume that all 3$d$-orbitals have 
the same radial wave-function (`spherical approximation'), 
all parameters are determined by, e.g., the 
three Racah parameters $A,B,C$. 
For comparison with other work, we 
introduce the average Coulomb interaction between electrons in the
same 3$d$-orbitals, $U=\sum_{c}U(c,c)/5=A+4B+3C$,
the average Coulomb interaction 
between electrons in different
orbitals, $U'=\sum_{c<c'}U(c,c')/10=A-B+C$,
and the average Hund's-rule exchange interaction,
$J=\sum_{c<c'}J(c,c')/10=5B/2+C$ that are related by
the symmetry relation $U'=U-2J$, see \ref{app:UJ}.
Due to this symmetry relation, the three values of $U$, $U'$, and $J$ do not 
determine the Racah parameters $A,B,C$ uniquely. Therefore, we make use
of the relation $C/B=4$ which is a reasonable assumption
for metallic nickel~\cite{buenemann2005,Sugano1970}.
In this way, the three Racah parameters and, 
consequently, all parameters 
in $\hat{V}^{\rm full}_{\rm loc}$ are functions of~$U$ and~$J$,
$A=U-32J/13$, $B=2J/13$, $C=8J/13$.
This permits a meaningful comparison of our results for 
all local Hamiltonians. Later we shall compare our results for 
$\hat{V}^{\rm dens}_{\rm loc}$ 
with orbital-independent values for $U$, $U'$ and $J=(U-U')/2$,
see eq.~(\ref{jo1}), with those for the full local Hamiltonian,
$\hat{V}^{\rm full}_{\rm loc}$, for the same values for $U$ and $J$.

\subsubsection{Double counting corrections.}
There exists no systematic (let alone rigorous) derivation of the double-counting 
correction in eq.~(\ref{eq:HubbardHamiltonian}). A widely used form for
this operator has first been 
introduced in the context of the LDA+$U$ method. Its expectation value
is given by 
\begin{equation}\label{78}
V_{\rm dc;1}^{\rm G}=\frac{U}{2}\bar{n}(\bar{n}-1)-
\frac{J}{2}\sum_{\sigma}\bar{n}_{\sigma}(1-\bar{n}_{\sigma})\;,
\end{equation}
where $\bar{n}_{\sigma}\equiv \sum_{c=1}^{N_c} C_{c,c;\sigma}^{\rm G}$,
$\bar{n}\equiv\bar{n}_{\uparrow}+\bar{n}_{\downarrow}$, and $N_c$ 
is the number of correlated orbitals ($N_c=5$ for nickel).
Note that  only the two mean 
values $U$ and $J$ enter this double-counting operator, i.e., it 
 is the same for all local Hamiltonians introduced above.
 
The physical consequences of the double-counting correction are most pronounced 
in its impact on the local energy-shifts $\eta_{c,c;\sigma}$ which we may write as 
 \begin{equation}
\label{78b}
\eta_{c,c;\sigma}\equiv \eta^{\rm G}_{c,c;\sigma} -\eta^{\rm dc}_{c,c;\sigma} \; ,
\quad \eta^{\rm dc}_{c,c;\sigma} 
=  \frac{\partial V_{\rm dc}^{\rm G}}{\partial C_{c,c;\sigma}}
\;. 
\end{equation}
For nickel, the cubic symmetry guarantees that
\begin{equation}
C^{\rm G}_{c,c';\sigma}=C_{c,c';\sigma}=\delta_{c,c'}C_{c,c;\sigma}
\; ,
\end{equation}
i.e., the correlated and uncorrelated local densities agree with each other.
The double-counting correction~(\ref{78}) 
leads to $\eta^{\rm dc,1}_{c,c;\sigma}=U(\bar{n}-1/2)+J(\bar{n}_{\sigma}-1/2)$. 
It has been argued 
in Ref.~\cite{dong2014} that this  double-counting correction is 
 insufficient for the investigation of cerium and some of its compounds. 
Instead, the authors of that work propose 
two alternative double-counting corrections which, in 
effect, correspond to the energy shifts 
\begin{eqnarray}
\label{78c}
V_{\rm dc;2}^{\rm G}: &&
\eta^{\rm dc,2}_{c,c;\sigma}=  \eta^{\rm G}_{c,c;\sigma}  \quad \hbox{so that 
$\eta_{c,c;\sigma}\equiv 0$,}
\\
\label{78d}
V_{\rm dc;3}^{\rm G}: &&
\eta^{\rm dc,3}_{c,c;\sigma}= \frac{1}{N_c} \sum_c\eta^{\rm G}_{c,c;\sigma} \;.
\end{eqnarray}
As we will demonstrate in the following section, 
these three double-counting corrections lead
to noticeably different results for nickel. 
This is a rather unsatisfactory observation because 
it compromises the predictive 
power of the method if the results strongly depend on 
the particular choice of the double-counting correction.
In Ref.~\cite{bloechel} a 
scheme has been proposed which does not rely on the subtraction of 
double-counting operators  and 
instead addresses the density functional directly. It needs to be seen
if this method can provide a more general way to tackle the double-counting 
problem within the Gutzwiller DFT. 

\subsection{Implementation in DFT}

We implemented our Gutzwiller scheme in the {\sc QuantumEspresso} DFT code;
for details on {\sc QuantumEspresso}, see Ref.~\cite{QuantumEspresso}.

Due to the cubic symmetry of nickel,
the single-particle density matrix $\tilde{C}$ is diagonal
with the local occupancies
$C_{t;\sigma}\equiv C_{\xi,\xi;\sigma}=C_{\eta,\eta;\sigma}=C_{\zeta,\zeta;\sigma}$
in the $t_{2g}$-orbitals and 
$C_{e;\sigma}\equiv C_{u,u;\sigma}=C_{v,v;\sigma}$
in the $e_{g}$-orbitals.
Likewise, the matrix $\tilde{\eta}$ is diagonal
with the corresponding entries $\eta_{t;\sigma}$ and $\eta_{e;\sigma}$.
Moreover, the $q$-matrix is diagonal, $q_{a,\sigma}^{b,\sigma}=\delta_{a,b}q_{a,\sigma}$
with identical entries for the three $t_{2g}$-orbitals,
$q_{\xi,\sigma}=q_{\eta,\sigma}=q_{\zeta,\sigma}\equiv q_{t,\sigma}$, and the two 
$e_g$-orbitals, 
$q_{u,\sigma}=q_{v,\sigma}\equiv q_{e,\sigma}$, respectively.
Formulae for $q_{t,\sigma}$ and $q_{e,\sigma}$ as a function of the 
Gutzwiller parameters $\tilde{\lambda}$ and of
$C_{e,;\sigma}$ and $C_{t;\sigma}$ are 
given in Refs.~\cite{buenemann2005,buenemann2012}.

\subsubsection{Setup: DFT calculation and Wannier orbitals.}

As a first step, we perform a DFT calculation that corresponds to setting
$U=J=0$. We use the LDA exchange-correlation potential,
\begin{equation}
v_{{\rm H,xc},\sigma}(\vecr)=
\left.
\frac{\partial
E_{\rm LDA,xc}\left[\left\{ n_{\sigma}(\vecr) \right\}\right]}%
{\partial n_{\sigma}(\vecr)} 
\right|_{n_{\sigma}(\vecr)=n_{\sigma}^0(\vecr)}
\; ,
\end{equation}
see eq.~(\ref{eq:EHxcisExcLDA}),
as implemented in {\sc QuantumEspresso}. The Kohn-Sham equations are solved
in the plane-wave basis using ultra-soft pseudo-potentials, see 
eq.~(\ref{eq:KSplanewaves}).
This calculation provides the Kohn-Sham bandstructure $\epsilon_{{\rm n},\sigma}(\veck)$
and the coefficients $C_{\vecG,{\rm n},\sigma}(\veck)=
\langle \veck, \vecG,\sigma|  \veck, {\rm n}, \sigma\rangle$
of the Kohn-Sham eigenstates $\psi_{\veck,{\rm n},\sigma}(\vecr)$
in the plane-wave basis.
The implemented `poor-man Wannier' program package provides 
the down-folded $3d$ Wannier orbitals $\phi_{\vecR,c,\sigma}(\vecr)$.
In the orbital Bloch basis the coefficients
$\langle \veck, \vecG, \sigma | \veck, c, \sigma\rangle$
describe $\phi_{\veck,c,\sigma}(\vecr)$ in the plane-wave basis.

\subsubsection{Gutzwiller--Kohn-Sham loop.}

At the beginning we set $q_{a,\sigma}^{b,\sigma}=\delta_{a,b}$ and $\tilde{\eta}=0$.
Our Gutzwiller--Kohn-Sham loop consists of the following steps.
\begin{enumerate}
\item \label{stepi}
Perform a DFT calculation with
the Gutzwiller Kohn-Sham 
Hamiltonian from eq.~(\ref{eq:Qetainplanewavebasis}), and 
the Gutzwiller Kohn-Sham densities from
eq.~(\ref{eq:densitiesorbital}).
Here, the form $Q_{\vecG,\vecG';\sigma}(\veck)
= \delta_{\vecG,\vecG'}+\sum_{c}(q_c -1)\langle \veck ,\vecG,\sigma | 
\veck,c,\sigma\rangle
\langle \veck,c,\sigma|  
\veck, \vecG', \sigma\rangle$ is useful where only the correlated orbitals
appear explicitly.

After reaching a self-consistent density $n_{\sigma}(\vecr)$,
calculate the local single-particle density matrix,
\begin{eqnarray}
C_{c,c;\sigma}&=& \frac{1}{L} \sum_{\veck,{\rm n}} f_{\veck,{\rm n},\sigma}\sum_{\vecG,\vecG'}
\langle 
\veck, {\rm n},\sigma | 
\veck, \vecG, \sigma
\rangle
\langle 
\veck, \vecG, \sigma | 
\veck, c, \sigma
\rangle
\nonumber \\
&&\hphantom{\frac{1}{L} \sum_{\veck} f_{\veck,{\rm n},\sigma}\sum_{\vecG,\vecG'}}
\langle 
\veck, c,\sigma | 
 \veck, \vecG', \sigma
\rangle
\langle 
\veck, \vecG', \sigma | 
\veck, {\rm n}, \sigma
\rangle \; ,
\end{eqnarray}
and the quantities $I^{\sigma}_{c_1,c_3,c_2,c_4}$ from eq.~(\ref{ki1}) and
$K^{\sigma}_{c,c'}$ from eq.~(\ref{ki2}).
For a proper convergence of the Gutzwiller--Kohn-Sham
loop these quantities must be calculated
with a momentum-space resolution that exceeds that of an ordinary DFT
calculation considerably. To achieve this goal we use a tetrahedron method with
826 $\veck$-points in the symmetry-reduced Brillouin zone.
\item \label{stepii}
Perform the inner minimization, i.e., minimize the energy 
functional $E^{\rm inner}$ in eq.~(\ref{dfhk}).
This step provides the values for the Lagrange parameters $\Lambda_l$ and
for the Gutzwiller variational parameters $\tilde{\lambda}$ 
that determine the orbital-dependent renormalization factors $q_{c,\sigma}$
in eq.~(\ref{eq:qfactorsdef}).
\item \label{stepiii}
Calculate the entries of $\tilde{\eta}$ 
from eq.~(\ref{eq:neededurgently}).
\item \label{stepiv} 
If the total energy does not decrease compared with the previous
iteration, the calculation has converged and the loop terminates.
If not, repeat the loop starting at step~(\ref{stepi}).
\end{enumerate}
The steps~(\ref{stepii}) and~(\ref{stepiii}) are carried out following 
the algorithm outlined previously~\cite{buenemann2012}.

In the present version of the program, step~(\ref{stepi})
requires a full DFT calculation which, however, is numerically cheap for
the simple nickel system. In the future, we plan to include the Gutzwiller minimization
directly in the DFT minimization cycle. 

The Gutzwiller approach permits the definition of
correlated orbital Bloch states, see \ref{subsubsec:pw-Gutzwiller}. 
Therefore, we can compare our original 3$d$ Wannier orbitals 
with the Gutzwiller correlated Wannier orbitals. For nickel, we find that
the deviations are negligibly small. In general, 
we may include the correlation-induced shape changes
of the correlated Wannier orbitals in our self-consistent 
calculations.

\begin{figure}[ht]
\begin{center}
\includegraphics[width=8.7cm]{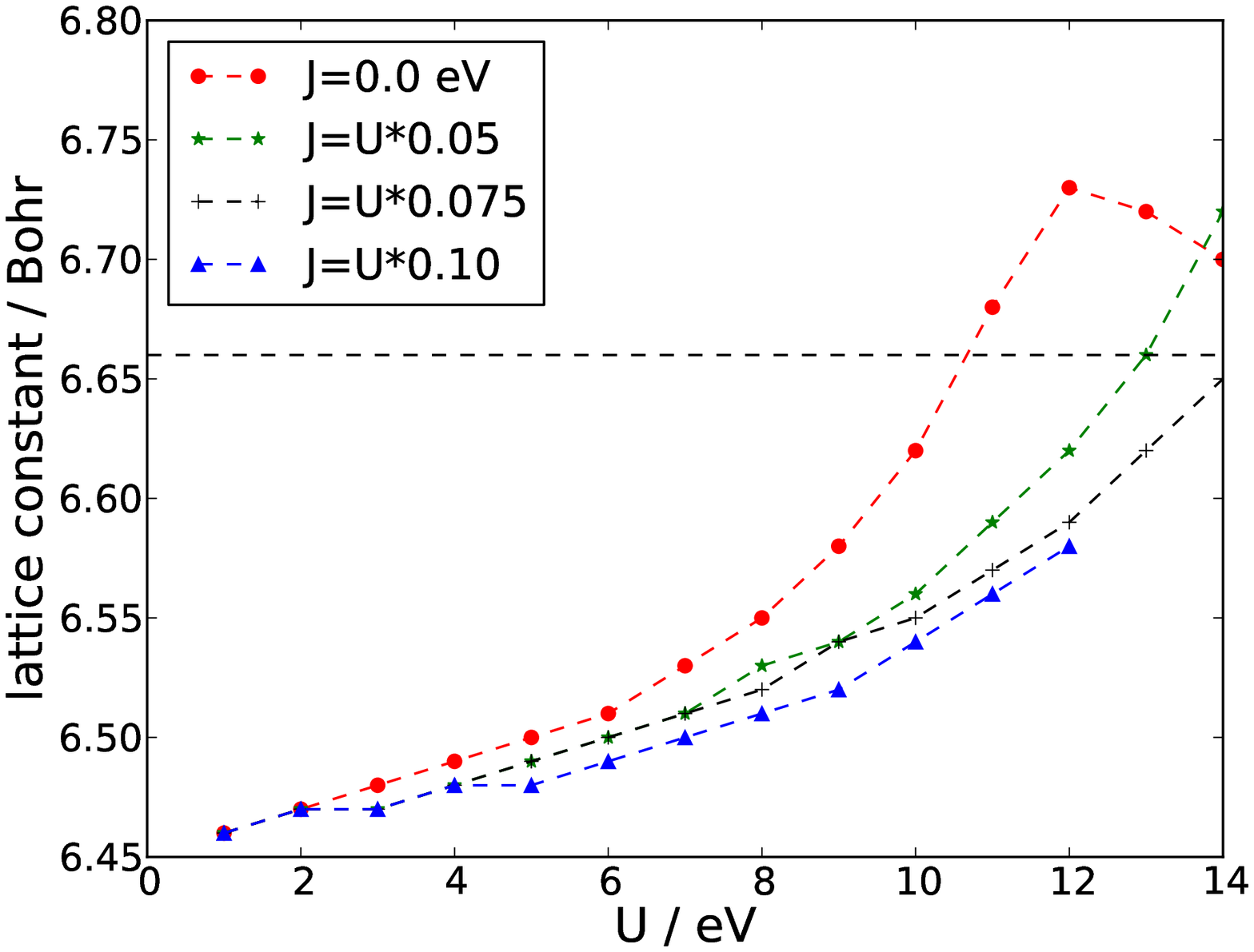}
\\
\includegraphics[width=8.7cm]{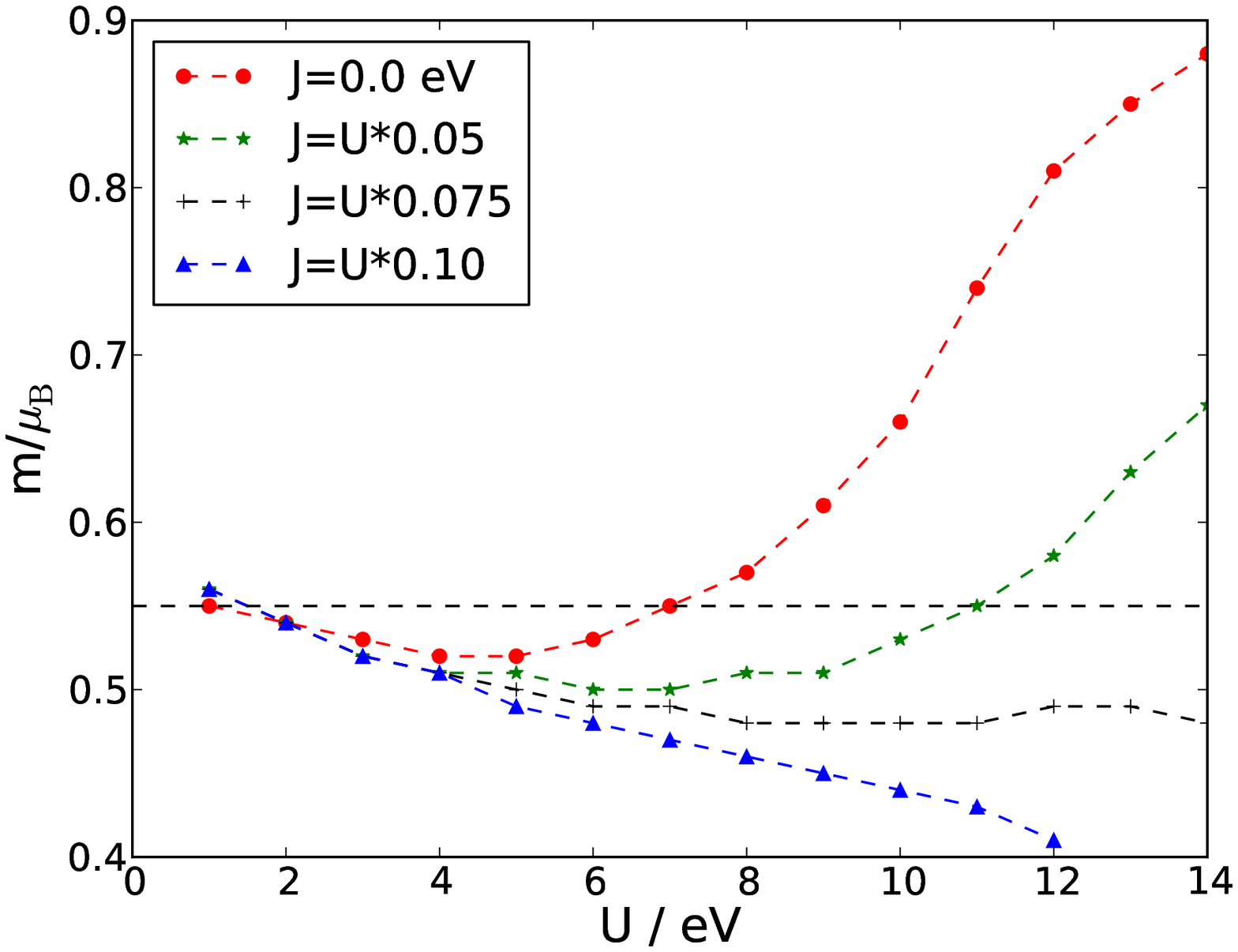}
\end{center}
\caption{Lattice constant (top) and magnetic moment (bottom)
of nickel as a function of $U$, for 
four different values of $J/U$, calculated with the full local Hamiltonian 
$\hat{V}^{\rm full}_{\rm loc}$ 
and the double counting correction $\hat{V}_{{\rm dc};1}$; dashed lines:
experimental values. \label{fig1}}      
\end{figure}

\subsection{Results}

The electronic properties of nickel 
have already been investigated by means of Gutzwiller 
wave  functions in Refs.~\cite{nickelpaper1,nickelpaper2,nickelpaper3}.  
In these works we started from a paramagnetic DFT-LDA calculation 
that provided the band parameters for a tight-binding model.
In order to overcome the deficiencies in the underlying DFT-LDA results, 
we fixed the magnetic moment and other single-particle properties 
at their experimental values.  As we will show in this section, 
the Gutzwiller DFT mends most of the DFT-LDA shortcomings. 
  
As a variational approach, the Gutzwiller DFT is expected to be most suitable for 
the calculation  of ground-state properties such as the lattice constant, 
the magnetic 
moment, or the Fermi surface of a Fermi liquid. 
Although more speculative than the ground-state calculations, it is also common to 
interpret  the eigenvalues of the Gutzwiller--Kohn-Sham Hamiltonian 
$\epsilon_{{\rm n},\sigma}^{\rm G}(\veck)$
as the dispersion of the single-particle excitations~\cite{Thul}. 
We shall discuss our results 
for the ground-state properties and single-particle excitations
separately.

\subsubsection{Lattice constant, magnetic moment, and bulk modulus of nickel.}

In Fig.~\ref{fig1}, we show the lattice constant and the
magnetic moment as a function of $U$ 
($1\, {\rm eV} \leq U \leq 14\, {\rm eV}$) for 
four different values of $J/U$ ($J/U=0,0.05,0.075,0.10$). 
In these calculations we used the full local Hamiltonian 
$\hat{V}^{\rm full}_{\rm loc}$ and the double-counting correction $\hat{V}_{{\rm dc};1}$. 

As is well known, the DFT-LDA underestimates the lattice constant. We obtain
$a_0^{\rm LDA}=6.47 a_{\rm B}$, considerably smaller than
the experimental value of $a_0=6.66 a_{\rm B}$ where 
$a_{\rm B}=0.529177\, \hbox{\AA}$ is the Bohr radius.
Fig.~\ref{fig1} shows that the Hubbard interaction~$U$ increases the lattice constant
whereby the Hund's-rule exchange~$J$ diminishes the slope. 
Apparently, a good agreement with the experimental lattice constant
requires substantial Hubbard interactions, $U> 10\, {\rm eV}$.

Fig.~\ref{fig1} shows the well-known fact that DFT-LDA reproduces
the experimental value for the spin-only magnetic moment $m_{\rm so}$ very well,
$m_{\rm so}^{\rm LDA}=0.58\mu_{\rm B}$ vs.\ $m_{\rm so}^{\rm exp}=0.55\mu_{\rm B}$.
However, when the DFT-LDA calculation is performed for the experimental
value of the lattice constant, the magnetic moment is grossly overestimated.
As seen in Fig.~\ref{fig1}, the Gutzwiller DFT allows us to reconcile the
experimental findings both for the lattice constant and the magnetic moment
if we work in the parameter range $11\, {\rm eV} < U < 14\, {\rm eV}$
and $0.05 < J/U < 0.07$. Note that a `fine-tuning' of parameters is not required 
to obtain a reasonable agreement between theory and experiment
for the lattice constant and spin-only magnetic moment. 

%
%
Our effective values are chosen to fit the experimental data for the 
lattice constant and the magnetic moment. The size of $U$ and $J$
agrees with those used in previous Gutzwiller-DFT studies 
on nickel~\cite{GutzwillerDFT2,nickelpaper1,nickelpaper2,nickelpaper3}. 
For the Gutzwiller-DFT the Hubbard-$U$ lies between
the bare, atomic value $U_{\rm bare}\approx 25\, {\rm eV}$~\cite{Czycholl}
and the low-frequency value for the screened on-site interaction 
$U_{\rm LDA+RPA}(\omega\to 0)\approx 4\, {\rm eV}$,
as obtained from LDA+Random-Phase Approximation~\cite{AryBiermann}
and used in LDA+DMFT~\cite{Biermann2}.
This comparison shows that the Gutzwiller-DFT works with a partly screened value
for the Hubbard interaction.
%
%

\begin{figure}[ht]
\begin{center}
\includegraphics[width=9cm]{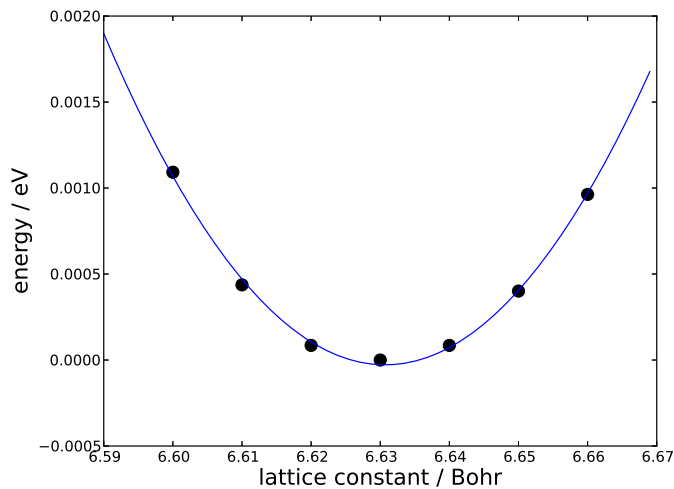}
\end{center}
\caption{Ground-state energy per particle $E_0(a)/N$
relative to its value at $a=6.63a_{\rm B}$
as a function of the fcc lattice parameter $a/a_{\rm B}$ in units of
the Bohr radius $a_{\rm B}$ for 
($U^{\rm opt}=13\, {\rm eV},J^{\rm opt}=0.9\, {\rm eV}$),
calculated with the full local Hamiltonian $\hat{V}^{\rm full}_{\rm loc}$ 
and the double counting correction $\hat{V}_{{\rm dc};1}$.
Full line: second-order polynomial fit.\label{fig2}}
\end{figure}         

For nickel, detailed information about the quasi-particle bands is available.
The quasi-particle dispersion at various high-symmetry points in the Brillouin zone is
more sensitive to the precise values of $U$ and $J$.
As we shall show below in more detail, we obtain a satisfactory agreement
with ARPES data for the choice 
$(U^{\rm opt}=13\, {\rm eV},J^{\rm opt}=0.9\, {\rm eV}$)
with an uncertainty of $\pm 1$ in the last digit.
For our optimal values we show in Fig.~\ref{fig2} 
the ground-state energy per particle $E(a)/N$ as a function 
of the fcc lattice constant~$a$ together with a second-order polynomial fit. 
The minimum is obtained at $a_0=6.63a_{\rm B}$, in good agreement
with the experimental value $a_0^{\rm exp}=6.66a_{\rm B}$.
For the magnetic spin-only moment we obtain $m_{\rm so}=0.52\mu_{\rm B}$, 
in good agreement with the experimental 
value $m_{\rm so}^{\rm exp}=0.55\mu_{\rm B}$.

{}From the curvature of $E(a)/N$ at $a=a_0$ 
we can extract the bulk modulus.
The bulk modulus at zero temperature 
is defined as the second-derivative of the ground-state energy
with respect to the volume,
\begin{equation}
K=V_0 \left. \frac{{\rm d}^2 E(V)}{ {\rm d} V^2 }\right|_{V=V_0}\; .
\end{equation}
This implies the Taylor expansion 
$E(V)=E(V_0)+(K V_0/2) (V/V_0-1)^2+\ldots $ for the ground-state
energy as a function of the volume $V=a^3$. For the ground-state energy
per particle we can thus write 
$E(a)/N=E(a_0)/N+e_2(a/a_{\rm B}-a_0/a_{\rm B})^2 +\ldots $
with 
\begin{equation}
e_2= \frac{9}{8} K a_{\rm B}^3 (a_0/a_{\rm B}) \; ,
\end{equation}
where we took into account that the fcc unit cell hosts four atoms, $V_0=N a_0^3/4$.
The fit leads to $K=169\, {\rm GPa}$, in good agreement with the experimental value,
$K=182\, {\rm GPa}$~\cite{bulkmodulusexp}.
It is  a well-known fact that the DFT-LDA overestimates
the bulk modulus of nickel. Indeed, 
our DFT-LDA gives $K^{\rm LDA}=245\, {\rm GPa}$. 

\begin{figure}[hbt]
\begin{center}
\includegraphics[width=8.8cm]{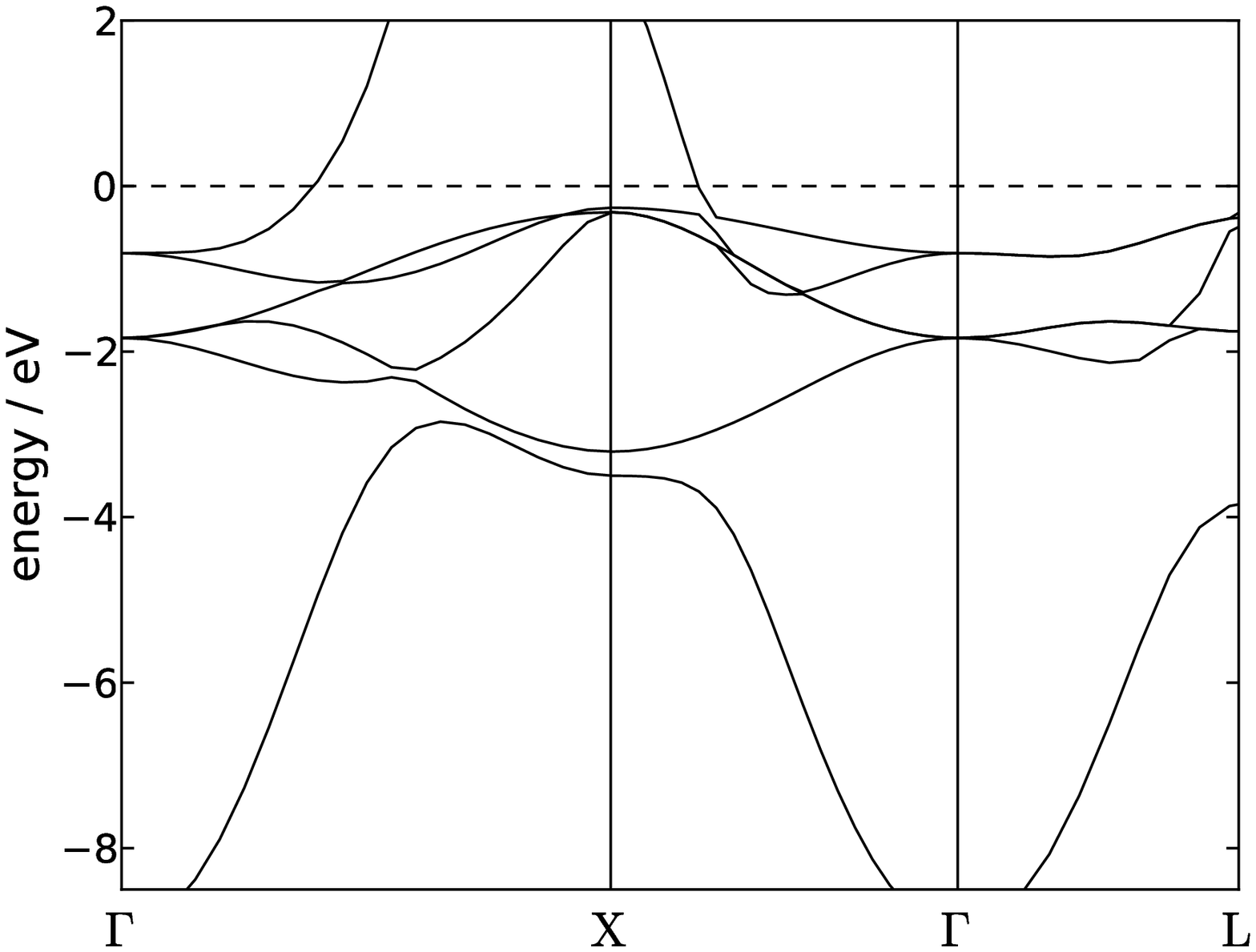}\\
\includegraphics[width=8.8cm]{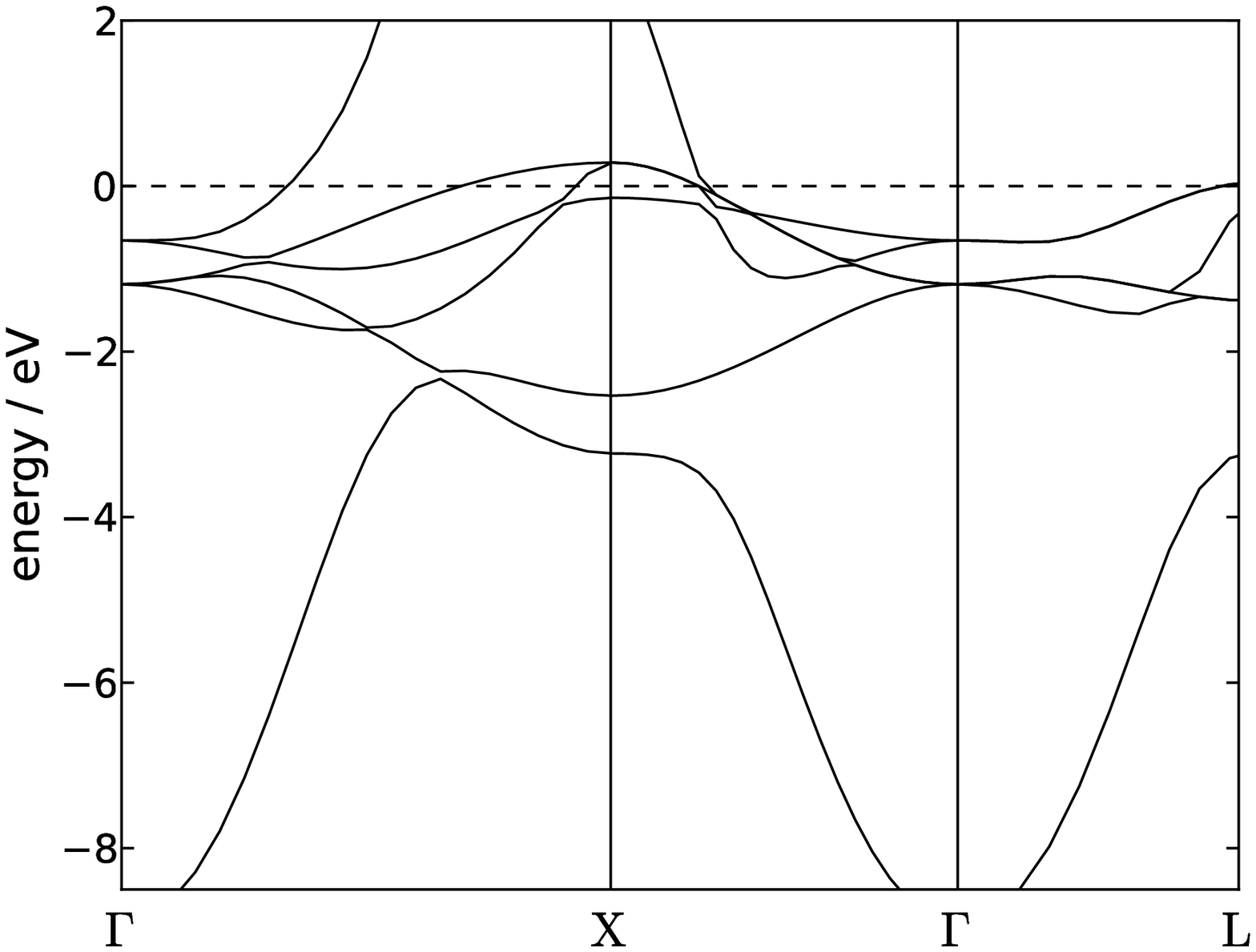}
\end{center}
\caption{Quasi-particle band structure of fcc nickel 
along high-symmetry lines in the first Brillouin zone,
calculated for $\hat{V}_{\rm loc}^{\rm full}$ and 
$\hat{V}_{{\rm dc};1}$ 
at  ($U^{\rm opt}=13\, {\rm eV},J^{\rm opt}=0.9\, {\rm eV}$).
Top: majority spin; Bottom: minority spin.
The Fermi energy is at $E_{\rm F}^{\rm G}=0$.
\label{fig:qpbands}}
\end{figure}         

We also calculated the lattice parameter and the magnetic spin-only moment
for the density-dependent interaction $V^{\rm dens}_{\rm loc}$, see eq.~(\ref{app3.5}),
with the same double-counting correction $\hat{V}_{{\rm dc};1}$.
Our results do not show significant discrepancies for the ground-state properties.
Note, however, that nickel is a special case 
because it has an almost filled 3$d$-shell ($n_{3d}\approx 9/10$) such that 
the terms from $\hat{V}_{\rm loc}^{\rm n.dens.}$
in eq.~(\ref{h255}) are more or less deactivated. Preliminary 
calculations for iron indicate that the missing interaction terms are more important 
for a partially filled $3d$-shell.

The combination of the full local 
interaction $V_{\rm loc}^{\rm full}$ with the second and third
double-counting correction,
see eqs.~(\ref{78c}) and~(\ref{78d}), 
does not lead to reasonable values for the lattice constant,
spin-only magnetic moment, and compressibility for nickel.
If we fix the lattice constant to its experimental value,
the Gutzwiller--Kohn-Sham equations lead to converged results for
the second (but not for the third) double-counting correction;
for the third double-counting correction, the $3d$~levels are discharged.
In the next section, we use these converged results for $\hat{V}_{{\rm dc};2}$ 
for comparison with those for the standard double-counting
correction $\hat{V}_{{\rm dc};1}$.

\begin{table}[t]
\begin{center}
\begin{tabular}{|c|c|c|c|c|}
\hline
\vphantom{\LARGE A}
Symmetry & Experiment & $\hat{V}_{\rm loc}^{\rm full}$ \& $\hat{V}_{{\rm dc};1}$ 
& $\hat{V}_{\rm loc}^{\rm full}$ \& $\hat{V}_{{\rm dc};2}$ 
 & $\hat{V}_{\rm loc}^{\rm dens}$ \& $\hat{V}_{\rm dc;1}$ \\[3pt]
\hline
$\langle \Gamma_1 \rangle$ & $8.90\pm 0.30$ & 8.95[0.08] & 8.99[0.08] & 8.93[0.08] \\ 
\hline
$\langle \Gamma_{25^{\prime}} \rangle$ & $1.30\pm 0.06$& 1.51[0.65] & 1.52[0.57] 
& 1.56[0.80] \\ 
\hline
$\langle \Gamma_{12} \rangle$ & $0.48\pm 0.08$ & 0.73[0.15] & 0.66[0.43] 
& 0.71[0.10] \\ 
\hline
$\langle X_1 \rangle$ & $3.30\pm 0.20$ & 3.37[0.27] & 3.26[0.56] & 3.42[0.10] \\ 
\hline
$\langle X_3 \rangle$ & $2.63\pm 0.10$ & 2.87[0.68] & 2.87[0.61] & 2.87[0.77] \\ 
\hline
$X_{2\uparrow}$ & $0.21\pm 0.03$ & \hphantom{$-$}0.26 & \hphantom{$-$}0.33 
& \hphantom{$-$}0.13 \\ 
\hline
$X_{2\downarrow}$ & $0.04\pm 0.03$ & \hphantom{$-$}0.14 & $-$0.06 & 
\hphantom{$-$}0.21 \\ 
\hline
$X_{5\uparrow}$ & $0.15\pm 0.03$ & \hphantom{$-$}0.32 & \hphantom{$-$}0.29 
& \hphantom{$-$}0.41 \\ 
\hline
$\Delta_{e_g}(X_2)$ & $0.17\pm 0.05$ & \hphantom{$-$}0.12 & \hphantom{$-$}0.39 
& $-$0.08 \\ 
\hline
$\Delta_{t_{2g}}(X_5)$ & $0.33\pm 0.04$ & \hphantom{$-$}0.60 
& \hphantom{$-$}0.51 & \hphantom{$-$}0.70 \\ 
\hline
$\langle L_1 \rangle$ & $3.66\pm 0.10$ & 3.49[0.61] & 3.49[0.56] & 3.55[0.83] \\ 
\hline
$\langle L_3 \rangle$ & $1.43\pm 0.07$ & 1.58[0.38]  & 1.52[0.52] & 1.61[0.26] \\ 
\hline
$L_{3\uparrow}$ & $0.18\pm 0.03$ & \hphantom{$-$}0.37 
& \hphantom{$-$}0.38 & \hphantom{$-$}0.34 \\ 
\hline
$\langle L_{2^{\prime}} \rangle$ & $1.00\pm 0.20$ & 0.14[0.06] & 0.17[0.06] & 0.12[0.06] \\ 
\hline
$\langle \Lambda_{3;1/2} \rangle$ & $0.50[0.21\pm 0.02]$ & 0.64[0.30] & 0.61[0.45] 
& 0.60[0.16] \\ 
\hline
\end{tabular}
\end{center}
\caption{Quasi-particle band energies 
with respect to the Fermi energy in eV 
at various high-symmetry points (counted positive for occupied states).
$\langle \ldots \rangle$ indicates the spin average, errors bars in the experiments 
without spin resolution are given as $\pm$. Theoretical data show the spin average 
and the exchange splittings in square brackets. $\Lambda_{3;1/2}$ denotes
the point half-way on the $\Lambda$-line that links the points $\Gamma$ 
and $L$. The first column gives experimental data
compiled in~\protect\cite{buenemann2005},
the second, third, and fourth column
give theoretical results results for 
$\hat{V}_{\rm loc}^{\rm full}$
with $\hat{V}_{{\rm dc};1}$,
$\hat{V}_{\rm loc}^{\rm full}$
with $\hat{V}_{{\rm dc};2}$,
and $\hat{V}_{\rm loc}^{\rm dens}$
with $\hat{V}_{{\rm dc};1}$, respectively, at
($U^{\rm opt}=13\, {\rm eV},J^{\rm opt}=0.9\, {\rm eV}$).
\label{tab:one}}
\end{table}

\subsubsection{Quasi-particle bands of nickel.}

In Fig.~\ref{fig:qpbands} we show the quasi-particle band structure
of fcc nickel for ($U^{\rm opt}=13\, {\rm eV},J^{\rm opt}=0.9\, {\rm eV}$).
The most prominent effect of the Gutzwiller correlator is 
the reduction of the $3d$ bandwidth. From a paramagnetic
DFT-LDA calculation one can deduce 
$W^{\rm LDA}=4.5\, {\rm eV}$~\cite{nickelpaper1,nickelpaper2,nickelpaper3}. 
whereas we find $W=3.3\, {\rm eV}$, in agreement with experiment.
This bandwidth reduction is due to the $q$-factors
$q_{t,\uparrow}=0.851$, $q_{t,\downarrow}=0.824$, 
$q_{e,\uparrow}=0.852$, $q_{e,\downarrow}=0.819$, 
$\bar{q}=\sum_{\sigma}(3q_{t,\sigma}+2q_{e,\sigma})/10=0.837$, so that 
$W\approx \bar{q}^2W^{\rm LDA}$.

A more detailed comparison of the quasi-particle band structure with experiment
is given in table~\ref{tab:one}. The overall agreement between experiment
and theory for $\hat{V}_{\rm loc}^{\rm full}$ with $\hat{V}_{{\rm dc};1}$
is quite satisfactory. In particular, only one hole ellipsoid is found
at the $X$-point, in agreement with experiment and in contrast to 
the DFT-LDA result~\cite{buenemann2005}.
Note, however, that the second double-counting correction 
$\hat{V}_{{\rm dc};2}$ spoils this advantage. Therefore, this form of the
double-correction term is not particularly useful for nickel.

We comment on two noticeable discrepancies between theory and experiment.
First, the energy of the band $L_{2'}$ at the $L$-point deviates by a factor of five.
This is an artifact that occurs already at the DFT-LDA level
and is not cured by the Gutzwiller approach.
Since the level has pure $3p$ character around the $L$ point,
the origin of the discrepancy is related to the
uncertainties in the partial charge densities $n_{3d}$, $n_{3p,3s}$ 
in the $3d$ and $3p/3s$ bands.
Second, the Gutzwiller DFT prediction for the 
exchange splitting $\Delta_{t_{2g}}(X_5)$ of the $t_{2g}$ bands at the $X$-point
is a factor of two larger than in experiment.
This deviation is related to the fact that, quite generally, all bands
are slightly too low in energy. This can be cured by decreasing~$U$ 
and increasing~$J$ but this deteriorates the values for the lattice
constant and the magnetic moment. 
We suspect that the deviations are partly due to the
use of a heuristic double-counting correction and the neglect
of the spin-orbit coupling.
Moreover, we expect the results for the band structure to improve
when we replace the `poor-man Wannier' orbitals
for the correlated $3d$ electrons by more sophisticated wave functions.

Table~\ref{tab:one} also shows the results for~$V^{\rm dens}_{\rm loc}$
with density-density interactions only and with $\hat{V}_{{\rm dc};1}$
as double-counting correction. The description provides the correct
Fermi surface topology but the deviations from the experimental band energies
is significantly larger. In particular, the
exchange splitting $\Delta_{e_{g}}(X_2)$ of the $e_{g}$ bands at the $X$-point
becomes negative, i.e., the order of the majority and minority bands is inverted.
The comparison of the band structures shows that 
the full atomic Hamiltonian should be used 
for a detailed description of the quasi-particle bands in nickel.

\section{Summary and conclusions}
In this work, we presented a detailed derivation of the Gutzwiller 
Density Functional Theory. 
Unlike previous studies, 
our formalism covers all conceivable cases of sym\-metries and 
 Gutzwiller wave functions. Moreover, our theory is not based 
 on the `Gutzwiller approximation'  which corresponds  to an 
 evaluation of expectation values in the limit 
of infinite lattice coordination number. It is only in the last step
 that we resort to this limit. 

 In particular, our derivation consists of three 
 main steps. 
\begin{itemize}
\item[1.] The density functional of the 
full many-particle system is related to that of a reference system with Hubbard-type
 local Coulomb interactions in the 
correlated orbitals. 
This generalizes the 
 widely used Kohn-Sham scheme where a single-particle reference system is 
 used. 
  \item[2.] The energy functional of the Hubbard-type reference system is 
 (approximately)
 evaluated by means of Gutzwiller variational wave functions. 
 \item[3.]  Analytical results for the energy functional are derived
  with the Gutzwiller approximation. 
\end{itemize} 
In a first application we studied the electronic properties of 
 ferromagnetic nickel. It turned out that the Gutzwiller DFT
 resolves the
main deficiencies 
  of DFT in describing ground-state properties such as the lattice 
 constant, the magnetic moment, or the  bulk modulus of nickel. 
Note that our approach requires  
  the relatively large value $U\approx 13\,{\rm eV}$ for the local 
 Coulomb interaction in order to obtain a good agreement with experiments. 
 
 Our results for the quasi-particle band structure 
 are by and large satisfactory. In fact, a perfect agreement 
 with ARPES data would be surprising because we calculate 
 these quantities based on Fermi-liquid assumptions that are strictly 
 valid only in the vicinity of the Fermi surface. Moreover, the 
  quasi-particle energies strongly depend on the orbital occupations
 that are influenced by the somewhat arbitrary choice of the
  double-counting corrections. As we have also shown in this work,
 different forms of the double-counting correction from  the literature 
 lead to fairly different results for nickel. 
 We consider this as the main shortcoming of the Gutzwiller DFT in its 
 present form that should be addressed in future studies. 
\label{sec:summary}


\appendix

\section{Single-particle systems}
\label{app:singleparticleproperties}

\subsection{Single-particle density matrix}
\label{app:showrhorhoequalsrho}

With the help of a single-particle basis $|k\rangle$
in which a given single-particle operator
$\hat{H}_{\rm sp}$ is diagonal, an eigenstate 
can be written as
\begin{equation}
|\Phi\rangle = \prod_{k}{}^{{}^{\prime}}
 \hat{b}_{k}^{\dagger} |{\rm vac}\rangle \; ,
\label{eq:Phifrombapp}
\end{equation}
where the prime indicates that $N$ single-particle states are occupied
in $|  \Phi\rangle$.
The single-particle density matrix is diagonal in $|\Phi\rangle$,
\begin{equation}
\rho_{k,k'}
\equiv
\langle \Phi | \hat{b}_{k}^{\dagger}
\hat{b}_{k'}^{\vphantom{\dagger}} | \Phi\rangle 
=
\delta_{k,k'}f_{k}\; ,
\end{equation}
and the entries on the diagonal obey
$f_{k}^2=f_{k}$ because we have $f_{k}=0,1$.
Therefore, we have shown that
\begin{equation}
\tilde{\rho}\cdot 
\tilde{\rho}
=\tilde{\rho} \; .
\label{eq:rhorhoapp}
\end{equation}
Since the operators $\hat{c}_{i}^{\dagger}$ for any other single-particle basis
and the operators $\hat{b}_{k}^{\dagger}$
are related via a unitary transformation,
eq.~(\ref{eq:rhorho}) holds generally for single-particle
density matrices for single-particle product states.

\subsection{Minimization with respect to the single-particle density matrix}
\label{appbueni}

We consider a general real function 
$E(\tilde{\rho})$ of a non-interacting 
density matrix $\tilde{\rho}$ with the elements
\begin{equation}
\rho_{i,j}=\langle \Phi |
\hat{c}_{j}^{\dagger} \hat{c}_{i}^{\vphantom{\dagger}}| \Phi \rangle\;.
\end{equation}
The fact that  $\tilde{\rho}$ is derived from a 
single-particle product  wave function $|\Phi\rangle$ is equivalent to the 
matrix equation~(\ref{eq:rhorho}).
Hence, the minimum of $E(\tilde{\rho})$ in the `space' of all  
non-interacting density matrices is determined by the condition
\begin{equation}
\frac{\partial}{\partial \rho_{j,i}}L(\tilde{\rho})=0\;,
\label{eq:variation}
\end{equation}
where we introduced the `Lagrange functional'
\begin{equation}
\label{sfg}
L(\tilde{\rho})\equiv 
E(\tilde{\rho})-\sum_{l,m}\Omega_{l,m}\Bigl(
\sum_p \rho_{m,p} \rho_{p,l}-\rho_{m,l}\Bigr)
\end{equation}
and the matrix $\widetilde{\Omega}$ 
of Lagrange parameters $\Omega_{l,m}$. 
Eq.~(\ref{eq:variation}) leads to the matrix equation
\begin{equation}
\tilde{H}=
\tilde{\rho}\cdot \widetilde{\Omega}
+\widetilde{\Omega}\cdot\tilde{\rho}
-\widetilde{\Omega}
\label{eq:matrixeqHappendix}
\end{equation}
for the `Hamilton matrix' $\tilde{H}$ with the elements 
\begin{equation}
H_{i,j}=
\frac{\partial}{\partial \rho_{j,i}}
 E(\tilde{\rho})\;.
\end{equation}
Equation~(\ref{eq:matrixeqHappendix}) 
is satisfied if eq.~(\ref{eq:rhorhoapp}) holds and if
\begin{equation}
[\tilde{H},\tilde{\rho}]=0\;.
\end{equation}
Hence, $\tilde{H}$ and $\tilde{\rho}$ must have the same basis 
of (single-particle) eigenvectors and, consequently, 
we find an extremum of $E(\tilde{\rho})$ if we choose
$|\Phi\rangle$ as an eigenstate of
 \begin{equation}
\hat{H}_{\rm sp}=\sum_{i,j}H_{i,j}
\hat{c}_i^{\dagger} \hat{c}_j^{\vphantom{\dagger}} \;.
\end{equation}
Usually, $|\Phi\rangle$ can be chosen as the ground state of $\hat{H}_{\rm sp}$.

\subsection{Basis sets}
\label{subsec:basissets}
\label{app:basissets}

\subsubsection{Kohn-Sham Hamiltonian in its eigenbasis.}

In the following we assume that the potential is lattice periodic,
\begin{equation}
V_{\sigma}^{\rm KS}(\vecr) = 
U(\vecr)+V_{\rm Har}(\vecr)+v_{{\rm sp,xc},\sigma}(\vecr)=
V_{\sigma}^{\rm KS}(\vecr+\vecR)
\; ,
\label{eq:defVsigma}
\end{equation}
where $\vecR$ is a lattice vector. The Fourier components are finite only
for reciprocal lattice vectors $\vecG$,
\begin{equation}
V_{\vecG,\sigma}^{\rm KS}= \frac{1}{V} 
\int {\rm d}\vecr V_{\sigma}^{\rm KS}(\vecr)e^{-{\rm i}\vecG\cdot\vecr} \; ,
\end{equation}
where $V$ is the crystal volume. As a consequence of the lattice periodicity,
the crystal momentum $\veck$ from the first Brillouin zone 
is a good quantum number.

As seen from eq.~(\ref{eq:definehKS}), 
the Kohn-Sham Hamiltonian
is diagonalized for the single-particle states
 $\psi_{{\veck},{\rm n},\sigma}(\vecr)=\langle \vecr | \veck, {\rm n},\sigma\rangle$ 
that obey 
\begin{equation}
h^{{\rm KS}}_{\sigma}(\vecr) \psi_{{\veck},{\rm n},\sigma}(\vecr)
=\epsilon_{{\rm n},\sigma} (\veck)
\psi_{{\veck},{\rm n},\sigma}(\vecr)\;,
\label{eq:KSequation}
\end{equation}
where $n$ is the band index.
Eqs.~(\ref{eq:KSequation}) are the Kohn-Sham 
equations~\cite{MartinLDAgeneral}.

In its eigenbasis, the Kohn-Sham Hamiltonian takes the form
\begin{equation}
\hat{H}^{\rm KS}=\sum_{{\veck},{\rm n},\sigma} 
\epsilon_{{\rm n},\sigma}(\veck)
 \hat{b}_{{\veck},{\rm n},\sigma}^{\dagger}
\hat{b}_{{\veck},{\rm n},\sigma}^{\vphantom{\dagger}} \; .
\end{equation}
Its ground state is given by 
\begin{equation}
|\Phi_0\rangle = \prod_{\sigma}\prod_{{\veck},{\rm n}}{}^{{}^{\prime}}
 \hat{b}_{{\veck},{\rm n},\sigma}^{\dagger}|{\rm vac}\rangle \; ,
\label{eq:Phifromb}
\end{equation}
where the $N$ levels lowest in energy are occupied as indicated by
the prime at the product, $\epsilon_{{\rm n},\sigma}(\veck)\leq E_{{\rm F},\sigma}$.
Then,
\begin{equation}
f_{{\veck},{\rm n},\sigma}=\langle \Phi_0 |  \hat{b}_{{\veck},{\rm n},\sigma}^{\dagger}
\hat{b}_{{\veck},{\rm n},\sigma}^{\vphantom{\dagger}} 
|\Phi_0 \rangle  =\Theta\left(E_{{\rm F},\sigma}-\epsilon_{{\rm n},\sigma}(\veck)\right)
\end{equation}
is unity for occupied levels up to the Fermi energy~$E_{{\rm F},\sigma}$,
and zero otherwise.

{}From eq.~(\ref{eq:expandpsi}), the field operators read
\begin{equation}
\hat{\Psi}_{\sigma}^{\vphantom{\dagger}}(\vecr)=\sum_{{\veck},{\rm n}} 
\psi_{{\veck},{\rm n},\sigma}(\vecr)
\hat{b}_{{\veck},{\rm n},\sigma}^{\vphantom{\dagger}}
\quad , \quad
\hat{\Psi}_{\sigma}^{\dagger}(\vecr) =\sum_{{\veck},{\rm n}} 
\psi_{{\veck},{\rm n},\sigma}^*(\vecr)
 \hat{b}_{{\veck},{\rm n},\sigma}^{\dagger}
\; . \label{eq:expandpsiKSbasis}
\end{equation}
Therefore, the ground-state density is readily obtained as
\begin{eqnarray}
n_{\sigma}^0(\vecr) &=&\langle \Phi_0 | \hat{\Psi}_{\sigma}^{\dagger}(\vecr)
\hat{\Psi}_{\sigma}^{\vphantom{\dagger}}(\vecr)| \Phi_0\rangle =
\sum_{{\veck},{\rm n}} f_{{\veck},{\rm n},\sigma} 
|\psi_{{\veck},{\rm n},\sigma}(\vecr)|^2
= \langle \vecr  | \sum_{\veck}  
\hat{\rho}_{\sigma}^{(0)}(\veck) 
| \vecr \rangle  \nonumber \; ,\\
\hat{\rho}_{\sigma}^{(0)}(\veck)  &=&
\sum_{n} f_{{\veck},{\rm n},\sigma}  | \veck ,{\rm n},\sigma\rangle
\langle \veck, {\rm n},\sigma |
\; ,
\end{eqnarray}
see also eq.~(\ref{eq:densitycond}).
Since this quantity enters the 
Kohn-Sham Hamiltonian, its solution
must be achieved self-consistently.

\subsubsection{Plane wave basis.}

In many codes, the Kohn-Sham Hamiltonian
is formulated in the plane-wave basis 
$|\veck, \vecG,\sigma\rangle$ with real-space representation
\begin{equation}
\langle \vecr | \veck, \vecG,\sigma\rangle =
\sqrt{\frac{1}{V}}e^{{\rm i} (\veck +\vecG)\cdot \vecr} \; .
\end{equation}
In this basis, the field operators are given by
\begin{equation}
\hat{\Psi}_{\sigma}^{\vphantom{\dagger}}(\vecr)=
\sqrt{\frac{1}{V}}
 \sum_{\veck,\vecG} 
e^{-{\rm i} (\veck +\vecG)\cdot \vecr} 
\hat{p}_{{\veck},\vecG,\sigma}^{\vphantom{\dagger}}
\; , \;
\hat{\Psi}_{\sigma}^{\dagger}(\vecr) =\sqrt{\frac{1}{V}}
 \sum_{\veck,\vecG} 
e^{{\rm i} (\veck +\vecG)\cdot \vecr} 
 \hat{p}_{{\veck},\vecG,\sigma}^{\dagger}\;,\label{eq:expandplanewaveKSbasis}
\end{equation}
and the Kohn-Sham Hamiltonian reads
\begin{equation}
\hat{H}^{\rm KS}= \sum_{\veck,\sigma} \sum_{\vecG,\vecG'}
T^{\rm KS}_{\vecG,\vecG';\sigma}(\veck)
\hat{p}_{\veck,\vecG,\sigma}^{\dagger}
\hat{p}_{\veck,\vecG',\sigma}^{\vphantom{\dagger}} \; .
\end{equation}
Eq.~(\ref{eq:defineTij}) shows that the entries of the
Kohn-Sham Hamiltonian in reciprocal space are given by
\begin{equation}
T^{\rm KS}_{\vecG,\vecG';\sigma}(\veck)=\delta_{\vecG,\vecG'}
\frac{1}{2m} \left(\veck +\vecG\right)^2
+ V_{\vecG-\vecG';\sigma}^{\rm KS} 
\end{equation}
for each $\veck$ from the first Brillouin zone. 
The eigenvalues of the Kohn-Sham matrix in reciprocal space
are $\epsilon_{{\rm n},\sigma}(\veck)$, and
the solution of the eigenvalue equation~\cite{MartinLDAgeneral}
\begin{equation}
\frac{1}{2m} \left(\veck +\vecG\right)^2C_{\vecG,{\rm n},\sigma}(\veck)
+\sum_{\vecG'} V_{\vecG-\vecG';\sigma}^{\rm KS} 
C_{\vecG',{\rm n},\sigma}(\veck) = \epsilon_{{\rm n},\sigma}(\veck)
C_{\vecG,{\rm n},\sigma}(\veck) 
\label{eq:KSplanewaves}
\end{equation}
for given $(\veck, {\rm n})$ gives
the entries of the eigenvectors,
$C_{\vecG,{\rm n},\sigma}(\veck)= 
\langle \veck,\vecG,\sigma| \veck,{\rm n},\sigma\rangle$.
Implemented plane-wave codes provide the band 
energies $\epsilon_{{\rm n},\sigma}(\veck)$
and the coefficients $C_{\vecG,{\rm n},\sigma}(\veck)$ so that the Kohn-Sham
eigenstates are obtained as
\begin{eqnarray}
|\veck,{\rm n},\sigma\rangle &=& 
\sum_{\vecG} C_{\vecG,{\rm n},\sigma}(\veck) |\veck,\vecG,\sigma\rangle \nonumber \; ,\\
\psi_{\veck,{\rm n},\sigma}(\vecr) &=& 
\sqrt{\frac{1}{V}}\sum_{\vecG} C_{\vecG,{\rm n},\sigma}(\veck) 
e^{{\rm i} (\veck +\vecG)\cdot \vecr}
\; .
\end{eqnarray}

\subsubsection{Orbital Wannier and Bloch basis.}
\label{app:orbialBlochbasis}

In order to make contact with many-particle approaches based on
Hubbard-type models, we need to identify orbitals that enter the
local two-particle interaction. 
Implemented plane-wave codes provide the transformation coefficients
$F_{(\veck,{\rm n}),(\vecR,b);\sigma}$ from Bloch eigenstates
$|\veck,{\rm n},\sigma\rangle$ to orbital
Wannier states $|\vecR,b,\sigma\rangle$,
\begin{equation}
|\vecR,b,\sigma\rangle 
= \sum_{\veck,n} F_{(\veck,{\rm n}),(\vecR,b);\sigma} 
|\veck, {\rm n},\sigma\rangle
\; , \;
F_{(\veck,{\rm n}),(\vecR,b);\sigma} =
\langle \veck,{\rm n},\sigma | \vecR,b,\sigma\rangle  
\; .
\end{equation}
The Wannier orbitals  
\begin{equation}
\phi_{\vecR,b,\sigma}(\vecr) 
=\langle \vecr |\vecR,b,\sigma\rangle
\end{equation}
are maximal around a lattice site $\vecR$ and 
the orbital index~$b$ resembles atomic quantum numbers, e.g., $b=3s,3p,3d$.
In the orbital Wannier basis the field operators are given by
\begin{equation}
\hat{\Psi}_{\sigma}^{\dagger}(\vecr) = 
 \sum_{\vecR,b} 
\phi_{\vecR,b,\sigma}^*(\vecr) \hat{c}_{\vecR,b,\sigma}^{\dagger} \quad, \quad
\hat{\Psi}_{\sigma}^{\vphantom{\dagger}}(\vecr) = 
\sum_{\vecR,b} 
\phi_{\vecR,b,\sigma}(\vecr) \hat{c}_{\vecR,b,\sigma}^{\vphantom{\dagger}}  \; , 
\label{eq:fieldoperatorsWannierbasis}
\end{equation}
and the Kohn-Sham Hamiltonian in the orbital Wannier basis becomes
\begin{equation}
\hat{H}^{\rm KS}= \sum_{\vecR,b,\vecR',b',\sigma}
T_{(\vecR,b),(\vecR',b');\sigma}^{\rm KS}
 \hat{c}_{\vecR,b,\sigma}^{\dagger} \hat{c}_{\vecR',b',\sigma}^{\vphantom{\dagger}} 
\end{equation}
with the overlap matrix elements
\begin{equation}
T_{(\vecR,b),(\vecR',b');\sigma}^{\rm KS}=
\int {\rm d}\vecr
\phi_{\vecR,b,\sigma}^*(\vecr)
h_{\sigma}^{\rm KS}(\vecr)
\phi_{\vecR',b',\sigma}(\vecr)
\; , \label{eq:defineTorbitalKS}
\end{equation}
see eq.~(\ref{eq:definehKS}).
These matrix elements appear in a tight-binding representation
of the kinetic energy in Hubbard-type models.

For later use we also define the orbital Bloch basis,
\begin{equation}
\phi_{\veck,b,\sigma} (\vecr)
= \sqrt{\frac{1}{L}} \!
\sum_{\vecR} e^{{\rm i}\veck \cdot \vecR}
\phi_{\vecR,b,\sigma} (\vecr) \,, \,
\phi_{\vecR,b,\sigma} (\vecr)
= \sqrt{\frac{1}{L}} \!\sum_{\veck} e^{-{\rm i}\veck \cdot \vecR}
\phi_{\veck,b,\sigma} (\vecr) \, ,
\end{equation}
where $\veck$ is from the first Brillouin zone and $L$ is the number of lattice sites. 
The field operators are given by
\begin{equation}
\hat{\Psi}_{\sigma}^{\dagger}(\vecr) = \sum_{\veck,b} 
 \phi_{\veck,b,\sigma}^*(\vecr) \hat{c}_{\veck,b,\sigma}^{\dagger} 
\quad , \quad 
\hat{\Psi}_{\sigma}^{\vphantom{\dagger}}(\vecr) = \sum_{\veck,b} 
 \phi_{\veck,b,\sigma}(\vecr) \hat{c}_{\veck,b,\sigma}^{\vphantom{\dagger}} 
\; . \label{eq:fieldoperatorsBlochorbiatlbasis}
\end{equation}
In the orbital Wannier basis, the Kohn-Sham single-particle Hamiltonian reads
\begin{eqnarray}
\hat{H}^{\rm KS}&=& \sum_{\veck,b,b',\sigma} T_{b,b';\sigma}^{\rm KS}(\veck)
 \hat{c}_{\veck,b,\sigma}^{\dagger} \hat{c}_{\veck,b',\sigma}^{\vphantom{\dagger}} 
\; , \nonumber \\
T_{b,b';\sigma}^{\rm KS}(\veck) &=&
\int {\rm d}\vecr
\phi_{\veck,b,\sigma}^*(\vecr) h_{\sigma}^{\rm KS}(\vecr)
\phi_{\veck,b',\sigma}(\vecr) \; . \label{eq:defineTorbital}
\end{eqnarray}

\section{Plane-wave basis for the Gutzwiller quasi-particle Hamiltonian}
\label{subsubsec:pw-Gutzwiller}

\subsection{Gutzwiller quasi-particle Hamiltonian in first quantization.}
\label{app:fristquant}
The Gutzwiller quasi-particle Hamiltonian in eq.~(\ref{eq:Hqpagain})
defines a single-particle problem in second quantization.
In order to express it in first quantization,
we define the single-particle operators
\begin{eqnarray}
\hat{\eta}_{\sigma}(\veck)&=&\sum_{b,b'} \eta_{b,b';\sigma} 
|\veck,b,\sigma\rangle\langle \veck,b',\sigma| 
\; ,\nonumber \\[3pt]
\hat{Q}_{\sigma}(\veck) &=& \sum_{a,b} q_{a,\sigma}^{b,\sigma}
|\veck,b,\sigma\rangle\langle \veck,a,\sigma|  \; .
\label{eq:etaQoperators}
\end{eqnarray}
The operator $\hat{\eta}_{\sigma}(\veck)$ is Hermitian. 

As seen from eqs.~(\ref{eq:Hqpagain}) and~(\ref{eq:TbbprimeG}), 
in the orbital Bloch basis we have
\begin{equation}
\hat{H}_{\rm qp}^{\rm G} = \!
\sum_{\veck,b,b',\sigma}
\langle \veck,b,\sigma | 
\left[
\hat{Q}_{\sigma}(\veck)
\hat{h}_{0,\sigma}(\veck) 
\hat{Q}_{\sigma}^{\dagger}(\veck)
+\hat{\eta}_{\sigma}(\veck)\right]
| \veck,b',\sigma \rangle
\hat{c}_{\veck,b,\sigma}^{\dagger}\hat{c}_{\veck,b',\sigma}^{\vphantom{\dagger}}
\label{eq:almostfirstquantized}, 
\end{equation}
where
\begin{eqnarray}
\hat{h}_{0,\sigma}(\veck) 
&=& \sum_{b,b'} 
h_{b,b';\sigma}^0 (\veck)
|\veck,b,\sigma\rangle\langle \veck,b',\sigma|  \nonumber \\
&=& \sum_{\vecG,\vecG'} h_{\vecG,\vecG';\sigma}^0(\veck) 
|\veck,\vecG,\sigma\rangle \langle \veck,\vecG',\sigma| \; .
\end{eqnarray}
Note that for the non-interacting limit, $\lambda_{\Gamma,\Gamma'}=1$,
we have $q_{a,\sigma}^{b,\sigma}=\delta_{a,b}$,
$\hat{Q}_{\sigma}(\veck)=\hat{1}$, and 
$\hat{\eta}_{\sigma}(\veck)=0$ so that
$\hat{H}_{\rm qp}^{\rm G}$ reduces to $\hat{H}^{\rm KS}$ 
in eq.~(\ref{eq:defineTorbital}).
Eq.~(\ref{eq:almostfirstquantized}) 
shows that the Gutzwiller quasi-particle Hamiltonian
in first quantization reads
\begin{equation}
\hat{h}_{\rm qp}^{\rm G}=\sum_{\veck,\sigma} 
\left[
\hat{Q}_{\sigma}(\veck)
\hat{h}_{\sigma}^0(\veck) 
\hat{Q}_{\sigma}^{\dagger}(\veck)
+\hat{\eta}_{\sigma}(\veck) \right] \; .
\end{equation}
This comparison proves relations used in previous
studies~\cite{GutzwillerDFT1,GutzwillerDFT2,dong2014}.

In the orbital Bloch basis
we define the operator
for the single-particle density matrix in first quantization as
\begin{equation}
\hat{\rho}_{\sigma}^{\rm G}(\veck) 
=\sum_{b,b'}\rho_{b,b';\sigma}^{\rm G}(\veck)
| \veck, b, \sigma\rangle \langle \veck, b',\sigma| \;,
\label{eq:densitiesorbital}
\end{equation}
with $\rho_{b,b';\sigma}^{\rm G}(\veck)$ from eq.~(\ref{eq:defnbbprimeGutzwiller})
where
\begin{equation}
\rho_{b,b';\sigma}(\veck)
= \langle \Phi_0 | 
\hat{c}_{\veck,b',\sigma}^{\dagger}
\hat{c}_{\veck,b,\sigma}^{\vphantom{\dagger}}
| \Phi_0\rangle 
=
\langle \veck, b, \sigma| \hat{\rho}_{\sigma}(\veck) | \veck, b', \sigma\rangle
\end{equation}
are the matrix elements 
for the optimized single-particle product state~$|\Phi_0\rangle$.
We define the projection operator $\hat{\rho}_{\sigma}(\veck)$ onto the occupied 
Gutzwiller quasi-particle states 
\begin{equation}
\hat{\rho}_{\sigma}(\veck)
=\sum_n f_{\veck,{\rm n},\sigma}^{\rm G} | \veck, {\rm n}, \sigma \rangle^{\rm G}\,
{}^{\rm G}\langle \veck , {\rm n},\sigma |
\; ,
\end{equation}
see eq.~(\ref{eq:qpdensities}).
With these definitions, we can readily express the local densities in 
eq.~(\ref{eq:Gutzdensities}) 
\begin{equation}
n_{\sigma}(\vecr)=\langle \vecr | 
\sum_{\veck} \hat{\rho}_{\sigma}^{\rm G}(\veck) 
| \vecr \rangle \; .
\label{eq:densitiesfirstquantization}
\end{equation}
Using the further assumption that the local single-particle density matrix
$\tilde{C}$ is diagonal and 
that $q_{a,\sigma}^{b,\sigma}=\delta_{a,b}q_{a,\sigma}$, 
we recover 
the expressions for the single-particle density matrix used 
in previous investigations~\cite{GutzwillerDFT1,GutzwillerDFT2}.

\subsection{Quasi-particle Hamiltonian 
in the plane-wave basis.}

Using the notation of \ref{app:fristquant},
we can readily express the Gutzwiller quasi-particle operator
in the plane-wave basis,
\begin{eqnarray}
\hat{h}_{\rm qp}^{\rm G} &=& \sum_{\veck,\vecG,\vecG',\sigma}
\langle \veck,\vecG,\sigma| 
\left[
\hat{Q}_{\sigma}(\veck)
\hat{h}_{\sigma}^0(\veck) 
\hat{Q}_{\sigma}^{\dagger}(\veck)
+\hat{\eta}_{\sigma}(\veck)\right]
| \veck ,\vecG',\sigma \rangle 
\hat{p}_{\veck,\vecG,\sigma}^{\dagger}\hat{p}_{\veck,\vecG',\sigma}^{\vphantom{\dagger}}
\;. \nonumber\\
\end{eqnarray}
This representation shows that
we have to diagonalize the Gutzwiller--Kohn-Sham plane-wave matrix with the entries
\begin{equation}
h^{\rm G}_{\vecG,\vecG';\sigma}(\veck)=
\sum_{\vecG_1,\vecG_2}\left[
Q_{\vecG,\vecG_1;\sigma}(\veck)
h_{\vecG_1,\vecG_2;\sigma}^0(\veck) 
Q_{\vecG',\vecG_2;\sigma}^{*}(\veck)\right]
+\hat{\eta}_{\vecG,\vecG';\sigma}(\veck)  \; ,
\label{eq:GQPfristq}
\end{equation}
where
\begin{eqnarray}
Q_{\vecG,\vecG';\sigma}(\veck)&=&
\sum_{a,b}q_{a,\sigma}^{b,\sigma}
\langle \veck ,\vecG,\sigma | 
\veck,b,\sigma\rangle
\langle \veck,a,\sigma|  
\veck, \vecG', \sigma\rangle \; ,\nonumber \\
h_{\vecG,\vecG';\sigma}^0(\veck)&=&\delta_{\vecG,\vecG'}
\frac{1}{2m} \left(\veck +\vecG\right)^2
+ V_{\vecG-\vecG',\sigma}^{\rm H} \; ,
\label{eq:Qetainplanewavebasis}\\
\eta_{\vecG,\vecG';\sigma}(\veck)
&= &
\sum_{b,b'} 
\eta_{b,b';\sigma} 
  \langle \veck ,\vecG, \sigma  
|\veck,b,\sigma\rangle
\langle\veck,b',\sigma|
\veck,\vecG',\sigma\rangle \; ,\nonumber 
\end{eqnarray}
for each $\veck$ from the first Brillouin zone. 
The eigenvalues of the Gutzwiller matrix 
are $\epsilon_{{\rm n},\sigma}^{\rm G}(\veck)$, and the entries
of the eigenvectors are $C_{\vecG,{\rm n},\sigma}^{\rm G}(\veck)$ so that
\begin{equation}
|\veck,{\rm n},\sigma\rangle ^{\rm G}=
\sum_{\vecG} C_{\vecG,{\rm n},\sigma}^{\rm G}(\veck) |\veck,\vecG,\sigma\rangle
\end{equation}
defines the Gutzwiller quasi-particle eigenstates in the plane-wave basis. From those
states we can derive `correlated orbital Bloch states'
$|\veck,c,\sigma\rangle^{\rm G}$ that can be used
to define the operators in eq.~(\ref{eq:etaQoperators}) self-consistently.
The correlations induce shape-changes
of the Wannier orbitals, i.e., $\phi_{\vecR,c,\sigma}^{\rm G}(\vecr)=
\langle \vecr | \vecR,c,\sigma\rangle^{\rm G}$ deviates from the original
Wannier orbital $\phi_{\vecR,c,\sigma}(\vecr)$.
Therefore, the correlated orbitals can be determined self-consistently.
We find that the effect is negligibly small for nickel.

\section{Atomic Hamiltonian in cubic symmetry}
\label{app:UJ}

We choose the Hubbard parameters
$U(u,v), U(\zeta,\zeta), U(\xi, \eta), U(\zeta, u), U(\zeta, v)$,
the four Hund's-rule couplings 
$J(u,v), J(\xi, \eta), J(\zeta, u), J(\zeta,v)$, 
and the two-particle transfer matrix element $S(\eta, \xi;\zeta, u)$
as our ten independent Coulomb matrix elements in cubic symmetry.
The other matrix elements in eq.~(\ref{h255})
can be expressed as~\cite{Sugano1970}
\begin{eqnarray}
\begin{array}{@{}lclcl@{}}
 U(u,u) &=& U(v,v) &=& U(u,v) + 2J(u,v)\;,\\ 
 U(\xi,u) &=& U(\eta,u) &=& (U(\zeta,u) + 3U(\zeta,v))/4\;, \\ 
 U(\xi,v) &=& U(\eta,v) &=& (3U(\zeta,u)+U(\zeta,v))/4\;, \\
 J(\xi,u) &=& J(\eta,u) &=& (J(\zeta,u)+3J(\zeta,v))/4\;, \\
 J(\xi,v) &=& J(\eta,v) &=& (3J(\zeta,u)+J(\zeta,v))/4\;, 
\end{array}\nonumber\\[6pt]
\begin{array}{@{}lclcl@{}}
 T(\eta;u,v) &=& - T(\xi;u,v) &=& \sqrt{3} (U(\zeta,u)-U(\zeta,v))/4\; ,\\
 A(\eta;u,v) &=& -A(\xi;u,v) &=& \sqrt{3} (J(\zeta,u)-J(\zeta,v))/4 \; ,
\end{array}\nonumber \\[6pt]
\begin{array}[b]{@{}lcl@{}}
 S(\xi,\eta;\zeta,u) &=& S(\eta,\xi;\zeta,u)\;, \\
 S(\zeta,\xi;\eta,u) &=& -2S(\eta,\xi;\zeta,u)\;, \\
 S(\xi,\eta;\zeta,v) &=& -\sqrt{3}S(\eta,\xi;\zeta,u)\;, \\
 S(\zeta,\xi;\eta,u) &=& \phantom{-} \sqrt{3}S(\eta,\xi;\zeta,u)\;.
\end{array}
\end{eqnarray}
If we further assume that the radial part of the $t_{2g}$-orbitals and the $e_g$-orbitals
are identical (`spherical approximation'), we may express all matrix elements
in terms of three parameters, e.g., the Racah parameters $A$, $B$, and $C$
that are related to the Slater-Condon parameters by $A=F^{(0)}-F^{(4)}/9$,
$B=(F^{(2)}-5F^{(4)}/9)/49$, and $C=5F^{(4)}/63$.
In particular, 
\begin{eqnarray}
U(u,v) &=& A - 4B + C\;,\nonumber \\
J(u,v) &=& 4B + C\;,\nonumber \\
U(\zeta,\zeta) &=& A +4B + 3C\;,\nonumber \\
U(\xi,\eta) &=& A - 2B + C\;,\nonumber \\
J(\xi,\eta) &=& 3B + C\;,\nonumber \\
U(\zeta,u) &=& A - 4B +C\;,\nonumber \\
U(\zeta,v) &=& A + 4B + C\;,\nonumber \\
J(\zeta,v) &=& C\;,\nonumber \\
J(\zeta,u) &=& 4B + C\;,\nonumber \\
S(\eta,\xi;\zeta,u) &=& -\sqrt{3}B\;.
\end{eqnarray}
The average Coulomb interaction between electrons in 
same orbitals is given by
\begin{equation}
U=\frac{1}{5}\sum_{c=\xi,\eta,\zeta,u,v}U(c,c)=
A+4B+3C \; ,
\end{equation}
the average Coulomb interaction between electrons in 
different orbitals is given by
\begin{equation}
U'=\frac{1}{10}\sum_{c, c'=\xi,\eta,\zeta,u,v;c<c'}U(c,c')=
A-B+C \; , 
\end{equation}
and the average Hund's-rule coupling becomes
\begin{equation}
J=\frac{1}{10}\sum_{c, c'=\xi,\eta,\zeta,u,v;c<c'}J(c,c')=
\frac{5}{2}B+C \; . 
\end{equation}


\begin{thebibliography}{99}
\bibitem{note1} 
The condition~(\ref{eq:DisDsp}) is too strong. 
We can equally work with 
$\widetilde{D}_{\rm sp}\left[\left\{ n_{\sigma}(\vecr) \right\}\right]
=D_{\rm sp}\left[\left\{ n_{\sigma}(\vecr) \right\}\right]+
\Delta\left[\left\{ n_{\sigma}(\vecr) \right\}\right]$
as long as  $\Delta\left[\left\{ n_{\sigma}^0(\vecr) \right\}\right]=0$
and $\left.(\partial \Delta\left[\left\{ n_{\sigma}(\vecr) \right\}\right])/
(\partial n_{\sigma}(\vecr))\right|_{n_{\sigma}(\vecr)=n_{\sigma}^0(\vecr)}
=0$. Then, $\widetilde{D}_{\rm sp}\left[\left\{ n_{\sigma}(\vecr) \right\}\right]$
leads to the same ground-state energy and ground-state density
as $D_{\rm sp}\left[\left\{ n_{\sigma}(\vecr) \right\}\right]$.
\end{thebibliography}
\end{document}